\documentclass[12pt,preprint]{aastex}

\shorttitle{WMAP and HI}
\shortauthors{Verschuur}

\begin{document}

\title{On the Apparent Associations between Interstellar Neutral Hydrogen Structure and ({\it WMAP}) High Frequency Continuum Emission}

\author{Gerrit L. Verschuur}
\affil{Physics Department, University of Memphis,
Memphis, TN 38152
gverschr@memphis.edu}

\begin{abstract}
Galactic neutral hydrogen (HI) within a few hundred parsecs of the Sun contains structure with an angular distribution that is similar to small-scale structure observed by the Wilkinson Microwave Anisotropy Probe ({\it WMAP}).  A total of 108 associated pairs of associated HI and {\it WMAP} features have now been cataloged using HI data mapped in 2 km/s intervals and these pairs show a typical offset of 0.\arcdeg8.  A large-scale statistical test for a direct association is carried out that casts little additional light on whether the these small offsets are merely coincidental or carry information. To pursue the issue further, the nature of several of the features within the foreground HI most closely associated with {\it WMAP} structure are examined in detail and it is shown that the cross-correlation coefficient for well-matched pairs of structures is of order unity.   It is shown that free-free emission from electrons in unresolved density enhancements in interstellar space could theoretically produce high-frequency radio continuum radiation at the levels observed by {\it WMAP} and that such emission will appear nearly flat across the {\t WMAP} frequency range.  Evidence for such structure in the interstellar medium already exists in the literature.  Until higher angular resolution observations of the high-frequency continuum emission structure as well as the apparently associated HI structure become available, it may be difficult to rule out the possibility that some if not all the small-scale structure usually attributed to the cosmic microwave background may have a galactic origin.

\end{abstract}

\keywords{Interstellar matter, neutral hydrogen, cosmology, WMAP}

\section{Introduction}
The interpretation of the distribution of the small-scale structure observed by the Wilkinson Microwave Anisotropy Probe ({\it WMAP}), as epitomized by the summary prepared from the five-frequency data in what is called the Internal Linear Combination ({\it ILC}) map, forms a cornerstone of modern cosmology.  The {\it ILC} map has been presented by Hinshaw et al. (2007) and has gone through several iterations, referred to there.  The key aspect of the {\it ILC} map is that the observed structure appears to be consistent with the existence of sound waves in the early moments of the universe, as summarized in the shape of the so-called acoustic spectrum.

It was disturbing to the present author to discover that some of the small-scale structure in the {\it ILC} data appeared to be closely associated with small-scale structure in the distribution of interstellar neutral hydrogen (HI) emission in the Galaxy, Verschuur (2007a, hereafter Paper 1, \& 2007b).  If these associations were to be regarded as anything other than mere coincidence they would imply that a previously unrecognized process occurring in interstellar space is capable of generating the structure in the high-frequency continuum emission (HFCE) observed by {\it WMAP}.  That, in turn, would imply that the cosmological interpretation of the {\it ILC} structures would be affected.

An immediate criticism of Paper 1 was produced by Land \& Slosar (2007) who searched for one-to-one correspondence between HI and {\it ILC} features.  However, that did little to either confirm or invalidate our earlier discussions, which were based on the discovery that peaks in the two forms of emission are closely associated but generally offset by a small angle.  This claim will be re-examined through a statistical test for a point-to-point correlation between the two forms of emission in narrow bands of HI emission.  When HI structure is mapped with a velocity resolution of 3 km/s at intervals of 2 km/s, this produces area maps at 200 different velocities between $-$200 to $+$200 km/s. (All velocities are given with respect to the local standard of rest.)  It is the apparent association between {\it ILC} structure and HI features in these velocity maps that reveals the possibility that there exist physical processes occurring in interstellar space capable of creating the observed HFCE.  The question that must then be considered is whether there is any possibility that such processes can occur in the interstellar medium.  It will be shown that unresolved structures containing excess electron density could, in principle, give rise to the HFCE at the levels observed by {\it WMAP}.  Evidence for structures of the required angular scale, particle density and temperature already exists in the literature.

The present analysis elaborates on the work reported in Paper 1 where it was pointed out that toward HI ÒcloudÓ MI there is a close association between {\it ILC} and HI structure as well as excess soft X-ray emission reported by Herbstmeier et al. (1995).  MI is one of two HI features at anomalous velocities discovered by Mathewson (1967) that came to be named MI and MII, the former at velocities around $-$110 km/s and the latter at about $-$80 km/s.  In the case of both MI and MII, associated H${\alpha}$ emission has been reported by Tufte, Reynolds \& Haffner (1998).  In Paper 1 it was predicted that a similar relationship between {\it ILC}, HI and X-ray structure found for MI would also be manifested in MII. The prediction is confirmed.

In \S3 the new statistical test is described.   In \S4 tables of HI-{\it ILC} associations in an area of sky below MII as well in directions of the ten brightest {\it ILC} peaks in an area of sky encompassing all galactic longitudes between $|${\it b}$|$ $=$ 30\arcdeg\ \& 70\arcdeg\ are presented. In \S5 the average angular separation between HI and associated {\it ILC} peaks found in Paper 1 and in the body of this paper are discussed. In \S6 the prediction made in Paper 1, that a relationship would be found between {\it ILC} structure, HI and soft X-ray emission for MII is discussed. The results of Gaussian analysis of the profiles toward HI features MI \& MII are outlined in \S7 in order to determine if something can be learned about the relationship of the narrow-band HI features to the {\it ILC} structure.  In \S8 a number of the very close associations between HI and {\it ILC} structure in the vicinity of both MI and MII are considered and in \S9 the results of cross-correlation calculations for several pairs of structures are presented. In \S10 it is shown that it is theoretically possible that unresolved  structure in the distribution of interstellar electrons can give rise the high-frequency emission at the same level as observed by {\it WMAP}.  In \S11 a variety of data pertaining to the presence of small-scale interstellar structure is considered in the light of the present work. Discussion is offered in \S12 followed by conclusions in \S13.

\section{Data}
The HI data used in this study were taken from the side-lobe-corrected, Leiden-Argentina-Bonn ({\it LAB}), All-Sky HI survey carried out with a beamwidth of 0.6\arcdeg\ and bandwidths 1.3 km/s (Kalberla et al. 2005).  These data were used to produce contour maps in galactic longitude and latitude (called {\it l,b} maps) of the total HI column density as well as the brightness temperature integrated over a 3 km/s velocity range plotted every 2 km/s in velocity over the relevant areas of interest.  During this study it became apparent that while a 10 km/s interval was used in Paper 1, a great deal more information emerges when the emission in a 3 km/s wide band plotted at 2 km/s intervals is used. In addition, position-velocity plots ({\it l,v} or {\it b,v} maps) were produced when needed and in several cases individual HI emission profiles were extracted from the LAB database for the purpose of Gaussian analysis to be described. 

In this report we will use the summary  {\it WMAP} data produced by Hinshaw et al. (2007) who combined the five-frequency data (in the range 23 to 94 GHz), after correcting for a number of effects including contributions from the galactic foreground, to produce the so-called Internal Linear Combination ({\it ILC}) map whose angular resolution was smoothed to 1\arcdeg.  Structure seen in this map is generally taken to represent conditions in the early universe.  

In order to understand the causes of the apparent associations between certain {\it ILC} and galactic HI features, observational studies should focus on those directions where the HI structure is relatively uncomplicated.  This rules out all directions close to the galactic disk.  Therefore the nature of the HI-{\it ILC} relationship in the galactic latitude band bordered by 30\arcdeg  and 70\arcdeg\ in both galactic hemispheres has been studied, the lower limit chosen to avoid the galactic disk, the upper limit set for practical purposes to avoid getting too close to the poles given that we use the rectangular galactic coordinate system of the LAB survey.  The bulk of the work reported here focusses on the northern galactic hemisphere data, particularly between {\it l} $=$ 160\arcdeg\ \& 210\arcdeg.  Examination of the data shown by Hinshaw et al. (2007) shows that this area of sky was subjected to the smallest corrections for foreground effects.

The first year {\it ILC} data made available to the author for the work initiated in Paper 1 were used for the present analysis and were supplemented by data for all longitude between latitudes 30\arcdeg\ \& 70\arcdeg\ for both galactic hemispheres.  Note that at the high galactic latitudes to which this work is confined there are no significant differences in the morphology of the {\it ILC} structure between the 1st and 3rd year {\it WMAP} results.  After all, very little if any correction for foreground continuum radiation (Hinshaw et al. 2007) was applied to the high galactic latitude data other than toward the spurs of galactic continuum emission, which lie outside the bounds of the regions considered in detail in this report.
     
\section{Large-scale statistical tests}
Given that nearly every (positive amplitude) {\it ILC} peak examined closely in Paper 1 and in what follows appears to be associated with an HI peak at some velocity, statistical tests to confirm or negate their significance may be moot.  The point is that no amount of statistical testing can prove a negative.  However, in order to determine if there is any evidence for a direct point-to-point correlation between the {\it ILC} and HI structure data, a simple statistical test was performed.  On the one hand the {\it ILC} data are available as intensities as a function of {\it l}  \& {\it b}.  Hence we have I$_{\it ILC}$ ({\it l,b}).  On the other hand the HI data are available as a function of velocity as well.  Thus the HI data are available in a data cube containing the brightness temperature, T$_{B}$ as a function of {\it l,b}, and {\it v}.  Hence T$_{B}$({\it l,b,v}).  The statistical test involves examining the product:  
\begin{equation}
P({\it v}) = T_{B}({\it l,b,v}).I_{\it ILC}({\it l,b})    
\end{equation}
at each of the velocities at which galactic emission HI over the area of interest is found.  This product was calculated using HI data integrated over a 3 km/s band (2 channels) at intervals of 2 km/s from $-$200 to $+$50 km/s over the northern sky between the latitude limits of 30\arcdeg\ \& 70\arcdeg\ and between $-$100 and $+$50 km/s for the southern sky data between the same latitude limits.  These limits were set by the extent of the HI emission in the two hemispheres.  Furthermore, the data were divided into thirds for areas in the northern (N) and southern (S) hemispheres as follows:

Target Area                {\it l} $=$ 60\arcdeg to 180\arcdeg\ (TN or TS)

Comparison Area 1  {\it l}  $=$ 300\arcdeg to 60\arcdeg\ (C1N or C1S)

Comparison Area 2 	{\it l}  $=$ 180\arcdeg to 300\arcdeg\ (C2N or C2S)

The appellation "Target Area"  grew out of the work reported in Paper 1 which focused on that longitude range in the north.

In addition to calculating the product, P({\it v}), for a given area, comparison calculations were performed by taking the HI data for that area, say TN, and then overlaying it on the {\it ILC} data for each of the two comparison areas, C1N and C2N, and repeating the calculations.  Thus three sets of products are calculated for each of the areas listed above:
\begin{equation}
P({\it v})_{TN.TN} = T_{B}({\it l,b,v})_{TN}.I_{\it ILC}({\it l,b})_{TN}
\end{equation}
\begin{equation}		
P({\it v})_{TN.C1N} = T_{B}({\it l,b,v})_{TN}.I_{\it ILC}({\it l,b})_{C1N}
\end{equation}
\begin{equation}			          
P({\it v})_{TN.C2N} = T_{B}({\it l,b,v})_{TN}.I_{\it ILC}({\it l,b})_{C2N}
\end{equation}
		
Similarly,  three sets of products were calculated for each of the Comparison Areas 1 \& 2.  The entire process was repeated for the data in the southern galactic hemisphere.  It is recognized that the HI and {\it ILC} data have different resolutions but that does not negate this search for apparent associations since the higher resolutions HI structures would appear to be encompassed by the lower resolution {\it ILC} data if the structures happened to overlap.
 
If the distribution of the {\it WMAP} structure and the HI peaks are random with respect to one another, a histogram of the product,
\begin{equation}  
P({\it v}) = T_{B}({\it l,b,v}).I_{\it ILC}({\it l,b}),
\end{equation}
should show a normal distribution.  If, however, there is a significant correlation between the two forms of emission within some velocity range the histogram should show a relative excess above a normal distribution at those velocities.  
 
Each group of three histograms covering 120\arcdeg\ in longitude and calculated every 2 km/s in velocity produced roughly 3 x 10$^{6}$ data points. For display purposes the results were plotted in contour map form for the number (N) of occurrences of the product with a given value (P) as a function of velocity.  Hence we display N(P,{\it v}).   For a given one third of the area studied (according to equations 2, 3 \& 4), three contour maps were plotted.  This produced a total of 9 such maps for the northern hemisphere and another 9 for the southern sky data.  In all, the full set of calculations involved 2.7 x 10$^{7}$  individual data points.
 
Two sets of values for the product were calculated for each data set.  One considered the value of P in steps of 0.05 K.mK and another over a limited range centered on zero in steps of 0.01 K.mK in case those revealed more detailed structure. 
 
In Paper 1 it was claimed that there exists a relationship between {\it ILC} structure with positive values and small-scale HI structure.  {\it ILC} structures that have positive values are deemed to represent directions of a small excess of high-frequency continuum radiation above the 2.738K cosmic microwave background and negative values represent slightly cooler areas, with the range from $-$0.22 to $+$0.22 mK. 
 
We have previously noted (and will discuss this further below) that there is a typical offset between HI and {\it WMAP ILC} peaks of order 0.\arcdeg8.  If this offset were larger than the typical width of the {\it ILC} and HI structures, no correlations should make their presence known in the map of the product, P(v).  If, instead, there is overlap between the contours defining structure in the two forms of emission the product, P(v), should reveal the presence of such associations in certain velocity ranges.
 
Figure 1 shows three histogram contour maps for the HI data for the Target Area in the north (TN) when correlated with the {\it ILC} data for Comparison Area 2 (at the left), with the {\it ILC} data for the Target Area (center), and when compared with the {\it ILC} data for Comparison Area 1 (right hand plot).  The product calculations were binned in steps of 0.01 K.mK in these plots.  Asymmetries toward negative product values implies a correlation between the HI structure and negative {\it ILC} structure.  These diagrams appear to reflect an excess of negative value signals in the {\it ILC} data in the northern strip of sky considered here.  Significant cases of direct positional associations between the HI and {\it ILC} data in the center plot should show up as marked asymmetries in the value of the product over a range of velocities pertaining to the relevant HI.  However, very little asymmetry is found in these plots bearing in mind that such small asymmetries as may be visible involve relatively few data points compared to the peaks in the histograms.  The individual histograms are themselves seldom simple Gaussians in shape.  This, too, is not surprising, since neither the distribution of large-scale distribution of HI and {\it ILC} structures over tens of degrees are not randomly distributed on the sky.  This is evident from a visual examination of the available data.  It is the small-scale associations that are of interest.  For example, the slight asymmetries around $-$50 km/s in the left-hand and right-hand plots indicate that in some areas of sky areas of excess HI structure happens to overlap areas of predominantly either positive or negative patches of {\it ILC} structure.  Thus taken together the data do not indicate significant overlap in position of the HI and {\it ILC} features at any favored velocity.
 
Land \& Slosar (2007) have performed a statistical test using HI data integrated in velocity bands 10 km/s wide over the whole sky and performed a search for direct correlations between {\it ILC} and HI structure in these velocity intervals.  However, any associations that might exist between the two forms of emission would not be expected to occur at a specific velocity over large areas of sky.  

The three plots in Fig. 1 and the other 15 similar plots that were produced for the northern and southern galactic hemispheres between $|${\it b}$|$ = 30\arcdeg\ \& 70\arcdeg\  show that convincing evidence for one-to-one associations is absent.  This does not contradict the presence of associations that manifest as small angular offsets between HI and {\it ILC} peaks.  Those are the ones that should be the focus of any further study.  The point is that close associations must first be identified by actually examining the relationship between the {\it ILC} and HI area maps and then calculating a cross-correlation coefficient on a case-by-case basis, taking into account that the associated peaks are usually offset by a small amount and that the axis joining the offset structures do not share a common position angle over the sky.   

\section{Newly Identified HI-{\it ILC} Pairs}
In this section a number of HI-{\it ILC} pairs will be cataloged in an area where the apparent association between the two types of structure is very obvious, even to the naked eye when examining the relevant HI and {\it ILC} data, for example the all-sky HI map found in the LAB web site (see astro.uni-bonn.de) and the widely publicized  {\it ILC} map (see map.gsfc.nasa.gov).  

\subsection{A Table of Associations in the MII Area}
HI feature MII is located at ({\it l,b}) $=$ (185\arcdeg, 63\arcdeg) and its relationship to {\it ILC} structure was studied in the context of a larger area bounded by {\it l} $=$ 180\arcdeg\ \&  210\arcdeg, {\it b} $=$ 50\arcdeg\ \&\ 70\arcdeg.  Figure 2a shows the total HI content for this area with the {\it ILC} contours overlain.  (In this and subsequent figures the HI data will be shown in color in the on-line version and as inverted gray-scale in the print edition.)  Figure 2b shows the same HI data in contour map form for visual comparison with Fig. 2a.  Obviously one cannot overlay two contour maps of this complexity in one figure and expect to learn anything by visual inspection alone.  The learning comes from taking the two contour maps and examining similarities and differences between them in a "blink comparison" mode. Some readers may prefer that such comparison be undertaken by computer, but one can only program the computer to find what one is looking for, and one cannot determine what one is looking for without first looking at the data! 

\subsection{A Trio of Associated Structures near MII}
In Fig. 2 only a few associations between HI and {\it ILC} peaks are obvious.   This impression changes dramatically when the HI data in 3 km/s wide bands are plotted at 2 km/s intervals.  For example, Fig. 3a shows a very bright (relative to its environment) HI feature at ({\it l,b}) $=$ (201.\arcdeg5, 56.\arcdeg5) in the velocity range $-$50 to $-$48 km/s which is all but invisible in the map of total column density (Fig. 2).  This HI peak is associated with an {\it ILC} peak observed by {\it WMAP} at ({\it l,b}) $=$ (201.\arcdeg2, 56.\arcdeg6) which is listed as \#79 in Table 1, below.  Figure 3b shows an elongated peak at ({\it l,b}) $=$ (203\arcdeg,55\arcdeg) found in the HI integrated between $-$40 \& $-$38 km/s.  Fig. 3c shows the HI brightness integrated between $+$1 and $+$3 km/s and it is associated with the prominent {\it ILC} peak at ({\it l,b}) $=$ (206.\arcdeg4, 55.\arcdeg0) listed as \#74 in Table 1.  In Fig. 3d the HI integrated brightness distribution in these three velocity ranges are plotted together and shows how closely the HI and {\it ILC} structures are correlated even if the HI is found at distinctly different velocities. Note that the slight offsets in longitude between the small angular diameter peaks in the HI and {\it ILC} data seen in Fig. 3 are ²0.\arcdeg3, or half the HI beam width.  Further details in the HI distribution in narrow velocity intervals will be discussed in \S6, 7 \&\ 8 below.

\subsection{A Table of Associations}
The 34 brightest peaks in the {\it ILC} contour map have amplitudes $>$0.063 mK and their positions and amplitudes are given in Table 1 together with the properties of the associated HI peaks found in the narrow-band maps.  Column 1 assigns an identification number for the {\it ILC} peaks in the MII region while continuing the numbering scheme begun in Paper 1, Table 1.  The longitude and latitude of each {\it ILC} peak is indicated in columns 2 \& 3 and the amplitude of the peak in mK in Column 4.  The center velocity of the 3 km/s wide band in which the associated HI feature was recognized is given in Column 5.  Columns 6 \& 7 give the longitude and latitude of the relevant HI feature and its peak amplitude in K.km/s in Column 8.  The angular offset in arcdegrees between associated pairs of peaks is given in Column 9.

In numerical terms 31 of the 34 {\it ILC} peaks in Table 1 reveal a closely spaced HI peak at some velocity when using these HI data plotted in 2 km/s intervals.  (This contrasts with only 7 of the 34 that appear to be associated in the HI map of total column density, Fig. 2.)  In all, 48 distinctly different HI features in the narrow-band HI maps appear to be associated with these 31 {\it ILC} peaks.  This is consistent with what was reported in Paper 1, which shows that in many cases HI at more than one velocity is apparently involved in the production of the HFCE, possibly at the interface between interacting HI features.  The average separation between paired set of HI-{\it ILC} peaks listed in Table 1 is 0.\arcdeg67 $\pm$ 0.\arcdeg35, which happens to be closely equal the beamwidth of the LAB HI survey.

The individual HI maps for this area between them contain 71 HI peaks defined as sets of closed contours that can be followed over adjacent velocity intervals, many of them showing small velocity gradients with changing position.  About 23 of these peaks, or 32\%, identified in the individual maps do not appear to have associated positive amplitude {\it ILC} peaks.  This compares with 6\% of the ILC peaks that do not show associated HI features.

A comparison was also carried out for the relationship between HI peaks and negative {\it ILC} ÒpeaksÓ with very similar results.  The area of Fig. 2 contains 30 distinct minima with amplitudes $<$ 0.056 mK and 18 of those appears to be associated with HI peaks.  The average separation is 0.\arcdeg90 $\pm$ 0.\arcdeg34.  Given the high density of structures in the {\it ILC} data for this region, with clear minima separating the peaks, it is to be expected that in general the minima would tend to mimic what is found for the maxima as regards their offsets with respect to HI peaks.  One way to become certain that the associations between the positive value {\it ILC} peaks and HI that are significant is to determine if they reveal anything new about interstellar physics.  Better still would be to obtain higher resolution observation of the HFCE and the HI structure to examine their apparent relationship more closely.

\subsection{The Ten Brightest {\it ILC} peaks}
The ten brightest {\it ILC} features in the area of sky for all galactic longitudes between {\it b} $=$ 30\arcdeg\ \&\ 70\arcdeg\ in both the northern and southern galactic hemisphere were also examined to determine if they were associated with HI structure.  All 10 sources, nine of which are found in the southern sky, are listed in Table 2 where the columns refer to the same properties as in Table 1.  The numbering in Column 1 continues from the numbering used in Table 1.  The velocities of the associated HI were again derived from examination of maps made at 2 km/s intervals.  Nine of the ten brightest {\it ILC} peaks can readily be related to the presence of HI and most of those involve HI peaks at more than one velocity, which is similar to what is found in Table 1.

\section{Average Separations of Associated Pairs}
The preliminary studies of possible associations between HI \& {\it ILC} structures discussed here and in Paper 1 so far includes 108 close pairs.  (An overview of the southern sky data for which we have prepared maps to continue this work reveal at least another 100 obvious close associations.  Details are deferred to a later paper.)  Table 3 summarizes the average angular separation between the {\it ILC} and HI peaks in each pair taken from three independent data sets considered so far.  The first entry is obtained from Table 1 in Paper 1.  The second entry is from Table 1 above and the third entry is from Table 2 above.  

The striking fact that emerges from the data in Table 3 is that the average apparent angular separations between HI peaks associated with {\it WMAP} {\it ILC} peaks are all of order 0.\arcdeg8.  This may be related to the method in which the associations were identified, which required that the HI contours defining a peak exhibit a morphology similar to the {\it ILC} feature for contours $>$0.02 mK.  For cases where the separations are larger than about 2\arcdeg, confusion quickly prevents an association from being recognized.  At that separation the {\it ILC} amplitudes tend to become negative.  
 
At this point we reiterate concerns that are raised by the overall situation summarized in Table 3. In total, of the 108 {\it ILC} peaks 102, or 94{\%}, show associated HI peaks.  The converse is not true in that not all HI peaks show assocciated {\it ILC} structure.  These cataloged associations involve HI at 152 distinct center velocities, which suggests that it could be the interaction between HI at different velocities that gives rise to the weak HFCE observed by {\it WMAP}.  Given that the HI is pervasive and extremely patchy, this abundance of associations is not surprising and could obviously be fortuitous.  The issue then becomes one of deciding whether it is possible that a previously unrecognized process in interstellar space could produce the HFCE observed by {\it WMAP} (see \S10).  

\section{A Close Look at the HI-{\it ILC} Association in the MI \& MII Areas}
In Paper 1, Fig. 6 the relationship between {\it ILC}, HI and excess soft-ray emission was published for MI.  Those data are reproduced here in Fig. 4a in which the shaded pixels representing the X-ray data taken from Herbstmeier et al. (1995) are plotted on a map of total HI content between velocities of $-$140 \& $-$80 km/s with the {\it ILC} contours overlain.  The nature of the HI emission underlying the shaded pixels will be discussed in \S7 below because here the brightest HI features are hidden by the X-ray pixel structure.  In Paper 1 it was predicted that something similar would be found for MII but the data that allowed that assertion to be tested only came to hand after that paper was submitted.  A similar plot for MII is shown in Fig. 4b where the HI data were integrated between $-$125 \& $-$55 km/s.  [Note that in these two areas very little HI emission is seen at low velocities, as reported by Verschuur \& Schmelz (2010a).]   Again it appears that all three forms of emission are related for MII and while in MI all three are closely associated in position for MII the peaks for each type of radiation are slightly offset from the other two.  If these close associations are due to anything other than pure chance then their relative location with respect to each other must surely provide information about the underlying physical processes involved in producing the HFCE and soft X-rays associated with the HI structure.

\section{Gaussian Analysis of HI Profiles Toward MI \& MII}
In order to more closely explore the possible relationship between HI and {\it ILC} structures revealed in the data, HI emission profiles in several areas of interest were decomposed into Gaussian components and the column densities of families of component line widths separately mapped on the sky.  This Gaussian decomposition was done by paying attention to the existence of underlying component 34 km/s wide that appears to be present in most if not all directions observed to date in the northern galactic hemisphere as reported in a preliminary study by Verschuur (2007b).  This broad component appears to be separately present in each of the so-called high-, intermediate-, and low-velocity regimes, hereafter referred to as HV, IV and LV.  This component is described by Verschuur \& Peratt (1999) and Verschuur (2004) and is readily identified in directions of the simplest HI profiles, while Verschuur (2007b) showed that in certain directions where the HI is least complex the pervasive nature of this component is most obvious. Therefore, in the present study a Gaussian component with this width was fit to each HI profile in each velocity regime while the Gaussian fitting algorithm (Verschuur, 2004) solved for the other components present. 

Note that it is possible that a Gaussian decomposition of a single profile may give ambiguous results.  However, when Gaussian analysis over an area is carried out, at least for relatively simple profiles, a remarkable coherence emerges when the results are considered as a whole.  Verschuur (2004) and references therein focused on relatively simple profiles to avoid ambiguity.  In this respect the profiles shown in Fig. 4 are in fact simple in that the three main peaks are well separated in velocity.  Thus each of those peaks is in itself simple, requiring at most three components to obtain a fit.  Severe problems related to ambiguity occur when four or more components closely blend, which is not the case for the areas mapped here.  A key fact that emerges from this analysis is that there is an underlying broad component of order 34 km/s present in each velocity regime.  Verschuur \& Schmelz (2010b) discuss the pervasive nature of this component which can clearly be seen in the profiles in Figs. 4a, e \& f.  It is also worth noting that Gaussian analysis of all the profiles in the LAB survey has been carried out by Haud and described in a number of papers, see for example Haud \& Kalberla (2007) and references therein, as well as by Nidever (private communication) who used an algorithm referred to in Nidever, Majewski \& Burton (2008).  Verschuur \& Schmelz (2010b) have used data from both Haud (private communication) and Nidever (private communication) to show that results for a set of common profiles agree very closely and that the pervasive component does not decompose into narrow lines when observed with a 9 \arcmin\ beam as opposed to the 36\arcmin\ beam of the LAB survey.  Thus the Gaussian mapping of the HI emission from MI (and MII, below) in the linewidth families discussed here is regarded as significant.

Figure 5 shows examples of the Gaussian decomposition for a number of directions toward MII where the presence of the broad underlying component is unambiguous.  After the Gaussian fitting was completed, a number of distinct linewidth and velocity families were recognized in the data.  These were sorted and maps made of the total column densities for each prominent category.    

The average parameters for the two dominant linewidth components associated with MI and MII are shown in Table 4.  Column 1 gives the name of the HI feature whose profiles were analyzed.  Column 2 gives the derived full-width, half-maximum linewidths in km/s averaged over the area under consideration with the 34 km/s value kept fixed.  Column 3 gives the average peak brightness temperature in K of the Gaussian components fit to the profiles sorted into line width family categories, Columns 4 gives the average center velocities in km/s with respect to the l.s.r. of these families, and Column 5 gives and average column densities in units of  10$^{18}$ cm$^{-2}$.

Table 4 shows that the Gaussian analysis revealed that HI emission profiles associated with MI and MII are dominated by two families of linewidths, those with this width of 34 km/s and another of order 21 km/s wide (ranging from 18 to 26 km/s). The common occurrence of Gaussian components of order 34 km/s wide has been discussed by Verschuur (2004) and references therein and the existence of components with widths of order 21 km/s wide have been noted by Haud (2008) as well as the present author in the case of anomalous velocity HI (in preparation).  It is possible that this component is due to HI at a temperature of 8,000 K and this will be discussed in a future paper.

\subsection{Gaussian Mapping for MI}
The dominant linewidth families found for MI have widths of 34 \& 21 km/s, Table 4, while two other families of linewidths have average widths of 13.8 $\pm$ 2.2 and 6.3 $\pm$ 2.4 km/s. Figure 6a shows the {\it ILC} contours superimposed on a map of the HI column density for the 34 km/s wide component associated with MI and Fig. 6b shows the relationship to the HI component with a linewidth of order 21 km/s.  In this figure the southern HI and {\it ILC} peaks overlap nearly perfectly while the 34 km/s wide component appears to be more prominently associated with the northern of the two {\it ILC} peaks.  Tufte, Reynolds \& Haffner (1998) have reported excess H${\alpha}$ emission at five locations toward MI and these are marked in Fig. 6b where it is obvious, as was also noted by those authors, that the peaks in the H${\alpha}$  emission are offset from the associated bright HI peaks.  When the two plots in Fig. 6 are compared with the map of total HI contents for MI (Fig. 4a above and  more specifically Fig. 6a in Paper 1) it is seen that it is the 21 km/s wide family of lines produces the main structure with the remaining contribution to the southern HI peak coming from the HI column density map in the two narrow components summarized in Table 4.  These narrower line width values are consistent with the data discussed by Verschuur \& Peratt (1999).  In Fig. 7 the column density maps for the two narrow component found in the emission profiles for MI.  Together these associations offer a clue as to how and why the two types of emission are related, provided one accepts the suggestion that these associations are due to something other than chance.  What the clues mean remains to be determined.

\subsection{A Closer Look at the HI and Soft X-ray Structure for MI}
The {\it ILC} peak in Fig. 4a at ({\it l,b}) $=$ (168\arcdeg,67.\arcdeg5) (listed as Source \#9 in Table 1 of Paper 1) has a small X-ray structure just to its south.  This is found to identically overlap an HI feature at positive velocities between $+$5 \& $+$10 km/s at ({\it l,b}) $=$ (169\arcdeg.67\arcdeg) shown in Fig. 8a.  In addition, an HI peak found at a velocity from $-$10 to $-$5 km/s shown in Fig. 8b closely abuts the HI component with a 21 km/s width seen in Fig. 6b and the relative morphology exhibits the same axial ratio for the elongated features.  These data strongly hint at complex interactions between HI features at different velocities interacting with one another to produce the HFCE observed by {\it WMAP}.  Also, if the low-velocity HI feature associated with the slight excess of soft X-ray emission at  ({\it l,b}) $=$ (169\arcdeg,67\arcdeg) is significant, then the association with the X-ray structure for the bulk of MI is not related to the mere presence of high-velocity gas. 

\subsection{Maps of HI components towards MII}
In the area toward MII bounded by {\it l} $=$ 180\arcdeg\ \&\ 194\arcdeg, {\it b} $=$ 62\arcdeg \&\ 68\arcdeg, 225 HI profiles were decomposed into Gaussian components in the same way as was done for MI.  Figure 9a shows the {\it ILC} contours overlain on a map of the HI column density in the 34 km/s wide component associated with MII, and Fig. 9b shows the same contours compared to the component with an HI linewidth of order 21 km/s.  Tufte, Reynolds \& Haffner (1998) have reported excess H${\alpha}$ emission at two locations toward MII.  These are shown in Fig. 9b.  Their direction \#2b at ({\it l,b}) $=$ (186\arcdeg,65\arcdeg) lies on the peak in the HI map of the 21 km/s wide feature.  The velocities of the HI component and H${\alpha}$  data are nearly identical at this position, $-$81.1 km/s for the HI and $-$72 or $-$78 km/s for each of two estimates by Tufte et al. (1998) for the H${\alpha}$ data.  The underlying 34 km/s wide component at this position is centered at $-$82.3 km/s.   

Both of the two HI components mapped in Fig. 9 are slightly offset from the {\it ILC} peak.  This is also true of excess soft X-ray emission associated with MII as reported by Herbstmeier et al. (1995) shown in Fig. 4b.  There is one bright {\it ILC} peak at ({\it l,b}) $=$ (193.\arcdeg5,63.\arcdeg8) at the edge of the maps in Fig. 4b.  It is Source \#68 in Table 1.  In Figure 10 the column density of a positive velocity component in the range $+$3 to $+$10 km/s is compared to the {\it ILC} contours and it appears that that {\it ILC} peak is associated with an excess of positive velocity HI in this direction.  This plot also shows that positive velocity HI emission appears to encompass the southern half of {\it ILC} feature associated with MII.

\section{Associations in the Area Encompassing MI and MII}
In the course of our analysis it was found that the associations between HI and the {\it ILC} structures are most clearly revealed in the narrow band area maps made at 2 km/s intervals.  In Paper 1 a velocity range and interval of 10 km/s were used but in many directions this hides the relevant structures that cover an intrinsically smaller velocity range.  

\subsection{A Striking Association south of MI}
An area of sky just south of MI includes several examples of the variety of associations found when small-scale features in the HI and {\it ILC} data are compared.  Figure 11a shows the {\it ILC} contours overlain on the map of total HI content for an area bounded by {\it l} $=$ 180\arcdeg\ \& 165\arcdeg, and {\it b} $=$ 45\arcdeg\ \& 65\arcdeg. Many of the pairs of associated structure listed in Paper I are located in this area but few are revealed in this map of total HI column density. Fig 11b compares {\it ILC} structure and HI data integrated from $-$20 to $-$10 km/s in the same area.  At ({\it l,b}) $=$ (173\arcdeg,50\arcdeg) an association is visible that is listed as Source \#7 in Paper 1, Table 1 but it is barely recognized and is all but invisible in the map of total HI content in Fig. 11a.  It is shown in detail in Fig. 12a, which presents the brightness temperature for Source \#7 in a narrow-band close-up of its area.  

A Gaussian analysis was performed on 143 profiles located every 0.\arcdeg5 in latitude and longitude for the area bounded by {\it l} $=$ 176\arcdeg\ \& 170\arcdeg, {\it b} $=$ 48\arcdeg\ \& 53\arcdeg\ and again an underlying component 34 km/s wide was readily identified in most profiles for both the low- and intermediate-velocity HI.  The most striking components that emerge after taking these into account is a set of narrow lines of order 3 to 5 km/s wide with a center velocity of $-$19 or $-$20 km/s over part of the area.  The HI column density of this component was mapped and the result is shown in Fig. 12b with the {\it ILC} contours overlain. A near perfect overlap, especially as revealed in the morphological boundaries is obvious. 

If the narrow width of this component is interpreted as a kinetic temperature, it is in the range of 180 K to 500 K.  At ({\it l,b}) $=$ (173\arcdeg, 50.\arcdeg5) the narrow HI component appears to be a double.  In Fig. 12b the column density of only one of these components was used to be consistent with the Gaussian solutions in its neighborhood but if the HI column density derived from the sum of these two components is used a map of the HI column density shown in Fig 12c is obtained.  The bright spot at ({\it l,b}) $=$ (172.\arcdeg, 50.\arcdeg5) implies that an additional HI structure is present there and this argues for obtaining higher angular resolution HI profiles for this area to better track the possible relationship to the {\it ILC} structure.  In general, the structure seen in Fig. 12 suggests that in this direction the source of HFCE is related to sources associated with an enhanced region of HI emission from cold components of the gas.

\subsection{Other associations near MI}
Figure 13 illustrates the relationship between the {\it ILC} and HI structure for four other pairs of features in the vicinity of MI. In Fig 13a a very close association between HI at $+$2 km/s and {\it ILC} source \#4 in Paper 1 is seen.  In Fig. 13b the HI brightness temperature at +2 km/s is shown overlain with the {\it ILC} contours for Source \#13 from Paper 1, located at ({\it l,b}) $=$ (169.\arcdeg5,46.\arcdeg5).  In that paper the positions were measured on a map made integrating over a 10 km/s range from $-$10 to 0 km/s and the offset in position between the peaks in the two forms of emission was estimated at 0.\arcdeg78.  In the data shown in Fig. 13b, the offset determined from a map made by integrating over the narrower velocity band of 3 km/s is zero.  The difference results from the presence of velocity broadening and velocity gradients within the structure seen in the HI.  Fig. 13b also reveals another pair of associated features near ({\it l,b}) $=$ (173\arcdeg, 46\arcdeg). This was identified as Source \#24 in Paper 1 but the data shown here indicate that this {\it ILC} peak is slightly shifted with respect to the position given there.  Figure 13c shows the HI at $-$136 km/s, which includes the {\it ILC} peak listed as source \# 40 in Paper 1 at ({\it l,b}) $=$ (167.\arcdeg3,55,\arcdeg7).   Clearly the HI at this velocity, which peaks at ({\it l,b}) $=$ (167.\arcdeg3, 55.\arcdeg0), is closely associated.  When the difference in angular resolution between the two types of data are taken into account it is visually evident that these structures are nearly identical in shape.  A Gaussian mapping of these areas was not undertaken other than to determine that in these directions the HI profiles showed components with line widths of 15 km/s for two of the peaks (from the left in Fig. 13), two overlapping components 15 and 4 km/s wide for the third peak, and two features 34 and 20 km/s wide for the right hand frame.  Again these hint at interacting gas masses being involved.

Fig. 14a compares the {\it ILC} contours with the HI data at $-$102 km/s integrated over a 2 km/s range.  {\it ILC} Source \#2 (from Paper 1) at ({\it l,b}) $=$ (174.\arcdeg4,56.\arcdeg8) closely abuts an HI feature at ({\it l,b}) $=$ (175.\arcdeg5, 55,\arcdeg0).  This association between the two forms of emission for Source \#2 is of particular interest since the boundary between the HI and the {\it ILC} peaks shows them to be unresolved at their interface.  This is illustrated in Figure 14b, which displays a cross-section of the amplitudes of both the HI and {\it ILC} peaks plotted along a line joining the center of the two peaks.  The equivalent half-widths at half-maximum intensity of the two features along this axis are 0.\arcdeg35 and 0.\arcdeg5 respectively, which compares with the beamwidths of 0.\arcdeg6 for the HI and 1\arcdeg\ for the {\it ILC} data.  These represent a case of abutting, unresolved edges.  The HI feature in Fig. 14a is centered at {\it b} $=$ 55.\arcdeg5 and has a total width of 0.\arcdeg64.  The {\it ILC} feature is intrinsically 1.\arcdeg2 wide, has a center at {\it b} $=$56.\arcdeg5.  The line of the interface is thus at b $=$ 55.\arcdeg8, which is coincident with the half-maximum height of both of the observed features.  These data suggest that the HFCE is being produced at the edge of the HI feature as projected on the sky.  In this case the Gaussian analysis of the peak HI profile shows overlapping components 20 and 15 km/s wide.

\subsection{The Trio of Structures}
The basic data for a trio of features in the same general area was shown in Fig. 3.  No attempts were made to map the Gaussian component structure for these directions, largely because the HI profiles are too complex for unambiguous mapping.  Gaussian analysis of the HI peaks in Fig. 3 showed that the relevant components had widths of  7 \&\ 4 km/s for the feature in (a) and 22 \&\ 4 km/s in (b) while the IV HI emission profile in the direction of the HI feature in (c) was too bright to be reliably decomposed into Gaussian components.

Overall, the data discussed in this section reinforce our contention that in order to carry out a comprehensive search for associations, use must be made of HI data plotted as narrow channel maps spaced at intervals of at most 2 km/s and preferably with high angular resolution so as to untangle the profile component structure, which so far has not shown any simple trends.  It is the HI morphology observed in small velocity intervals, and the velocity structure within in the profiles, that is likely to carry the most information regarding the nature of the complex physical interactions that appear to underlie the production of HFCE in interstellar space. 
 
\section{Cross correlation studies}
Given that there appear be small angular offsets between specific {\it ILC} and HI features as discussed in \S7 \& \S 8, it is worth examining whether there is a statistically significant relationship when they are considered on a case-by-case basis.  

In \S3 it was shown that there is little evidence for a direct one-to-one correspondence between the {\it ILC} and HI peaks over large areas of sky.  Furthermore, the close associations discussed so far do not occur in a uniform manner because the axes connecting the members of paired structures are not oriented in the same direction on the sky.  Hence no comprehensive, large-scale statistical test for a fixed position angle over large area of sky will reveal anything of significance.  However, a meaningful test could be to consider individual pairs of associated features and then shift one with respect to the other in position so as to align them even if they are initially slightly offset and then derive a cross-correlation coefficient to test for first-order significance.   In this regard, small offsets between the two forms of emission would be the hallmark of a form of limb brightening along one face of an HI feature, where the neutral gas interacts with the surrounding interstellar plasma, to be discussed below.

Table 5 shows the results of the correlation calculations for some of what are mostly unresolved pairs of structures discussed above and in Paper 1.  Column 1 gives the source name and/or references used in this work and a reference to the figure number displaying the data.  Column 2 gives the velocity of the HI at which an association is claimed while columns 3\ \&\ 4 give the magnitude of the cross-correlation coefficient when the two data sets are compared along cross sections in longitude \& latitude.  The values obtained using the data un-shifted in position are shown followed by the values obtained when the calculation is carried out after shifting the HI data with respect to the {\it ILC} data by the angle indicated in order to align the peaks. Note that if two unresolved features observed with the beam widths of 0.\arcdeg6 and 1.\arcdeg0 are perfectly coincident in position the cross-correlation coefficient would be 0.93.  Thus those cases listed in Table 5 that show a value of R of this magnitude are in fact highly correlated, which means that the members of each associated pair have virtually the same angular dimensions.  It could be argued that the method of identification of the closely associated pairs in the first place favors those that look closely similar.  For example, one would tend to avoid attaching significance to finding an {\it ILC} feature of large angular extent next to a small HI feature, or vice versa.  However, during the discovery of the associations it was not obvious that this was occurring.

\section{A Possible Model}
In this section it is assumed that the spatially offset associations of {\it ILC} and HI features is significant and will consider whether free-free emission from unresolved structures in the distribution of electrons in interstellar space could give rise to HFCE.  Note that contrasts with the case of free-free emission from electron-ion pairs referred to by Hinshaw et al. (2007).

\subsection{Free-free Emission from Electrons}
Using the form for free-free emission as given, for example, by Nitta et al. (1991), the optical depth $\tau_{ff}$ can be expressed as a function of frequency, $\nu$,  the electron temperature, $T_e$, the electron density, $n_e$, and path length, l, as follows:

\begin{equation}
\tau_{ff} = 9.8 \times 10^{-3}\  \nu^{-2}\  T^{-1.5}_e\  ln (4.7 \times 10^{10}\ T_e/\nu) \int n_e^2\  dl.
\end{equation}

For low optical depth the observed brightness temperature as a function of $\nu$ is then:

\begin{equation}
T_B(\nu) =  \tau_{ff} \ T_e.
\end{equation}

The expected continuum signal expected from free-free emission as defined by  Eqtns. 6 \& 7 can be calculated from the angular width of the electron enhancements (or cloud) on the sky, $\theta$$_o$, and the Aspect Ratio (depth of feature relative to its width) for a given distance as well as the derived linear diameter or width $D_o$.

For a distance L in pc and expressing the scales in cm, the linear diameter is given by
\begin{equation}
D_o =  5.3\times 10^{16}\  \theta_o\ L\  cm
\end{equation}

If the linear depth of the electron cloud in the beam that produces the observed column density is defined as D$_l$ then,
\begin{equation}
D_l =  5.3\times 10^{16}\  \theta_o\ A\ L\  cm
\end{equation}

If $N_H$\ is the observed HI column density in units of 10$^{18}$ cm$^{-2}$, then the volume density, $n_H$, is given by:
 
 \begin{equation}
n_H = 19\ N_H\ (\theta_o\ A\ L)^{-1}\  cm^{-3}.
 \end{equation}
 
For a fraction, {\it f}, of the HI density in the form of electrons, the electron column density $N_e$ allows the electron volume density, $n_e$, to be derived in order to determine the emission measure.  Here it is assumed that in general the electron "clouds" are associated with the adjacent HI features of known column density, although the two could be quite unrelated depending on how the electrons are produced.

Using the above we find that
\begin {equation}
n_e = 18.9\ f\  N_H\ (\theta_o\ A\ L)^{-1}\  cm^{-3}
\end{equation}

Using Eqtn. (11) and substituting in Eqtns. (6) \& (7), the brightness temperature as a function of frequency $\nu$\ and the observed HI column density is then given by:

\begin{equation}
T_B(\nu) = 1.86\times10^{17}\  \nu^{-2}\ T^{-0.5}_e\ ln (4.7 \times 10^{10}\ T_e/\nu)\ \it f^2\ N_H^2\  (\theta_o\ A\ L)^{-1}\  K.
\end{equation}

Equation 12 assumes that the source fills the beam.  However, the observed brightness temperature will be diluted in the case of unresolved sources in the ratio of source area divided by the beam area. In evaluating the equation in the next section this will be considered.

\subsection{Evaluating the Equation:  Another Coincidence?}
In order to evaluate Eqtn. 12 information on distance, angular scale and temperatures of the electron features (clouds?) is required.  In the case of the most striking associations between HI and  {\it ILC}  structures shown in Paper 1 it was noted that in many cases HI at multiple velocities appears to be associated with a given  {\it ILC}  peak.  Thus the HI emission in three velocity regimes (high, intermediate- and low-velocity) is associated.  This suggests that all the relevant HI at high-latitudes is relatively local, within at most 200 pc of the Sun, if we use the canonical half-width of the galactic disk to be 100 pc and take into account the latitude of the areas under consideration.  This point is reinforced by the recent study of Verschuur \& Schmelz (2010a). Furthermore, the fact that HI at distinctly different velocities appears to be interacting in the area where the high-frequency emission is enhanced suggests that the phenomenon of interaction may be involved in triggering the creation of electrons at an interface between the interacting HI features, or where HI features interact with surrounding plasma through which they travel.  

Verschuur (1991) has argued that enhancements in HI brightness seen in twisted filamentary features will be produced where the filament geometry twists into and out of the line-of-sight to cause the total HI column density to increase in certain directions.  This defines an Aspect Ratio of A$>$1 for structure in filaments.  However, there is no {\it a priori} reason to reject the possibility that A$<$1, which would apply in the case of sheet-like structures along the line of sight.

A key parameter required to evaluate Eqtn. 12 is the electron temperature, $T_e$.  Two values will be tested below.  One corresponds to that of cold HI, approximately 50 K (e.g., Wakker et al. 1991 and a summary in Kulkarni \& Heiles 1988), and 8,000 K, corresponding to the temperature of ionized hydrogen.

The amplitude for free-free emission is expected to vary as $\nu^{-2}$ according to Eqtn. 12 with the dependence on $\nu$ encompassed by the {\it ln} term small in comparison. Hence between the two extremes of the {\it WMAP} frequency range of 23 \& 94 GHz the amplitude will decrease by a factor of nearly 16.  However, the beam widths are 0.\arcdeg88 \& 0.\arcdeg22 respectively and thus the beam dilution factor goes will increase the observed signal by a factor of 16 across the same frequency range, which effectively counterbalances the variation with frequency produced by free-free emission.  Thus the resulting spectrum produced by unresolved structure in the beam will, to first-order, appear flat across this frequency range.

\subsection{HFCE Produced by Unresolved Interstellar Electron Features}
In order to evaluate the possible magnitude of this signal expected from free-free emission from interstellar electrons, a first-order attempt to apply observable parameters to Eqtn. 12 taking into account beam dilution was undertaken.  The HI features most closely associated with {\it ILC} structures for MI and MII  discussed above have a typical column density of order 1.4 $\pm$ 0.3 x 10$^{20}$ cm$^{-2}$ with corresponding positive {\it ILC} average amplitude 0.16 $\pm$ 0.03 mK.  Clearly each pair of associated structures may be at a different distance, each may have a different Aspect Ratio, and each may have a different temperature. However, the purpose of this exercise is to determine whether it is possible, for a reasonable choice of parameters, to expect that HFCE at the level of 0.16 mK could be produced by free-free emission from clumps of interstellar electrons.  

Figure 15 plots the expected brightness temperature at both 23 and 94 GHz as a function of total electron column density for a number of models.  Their properties were chosen to be illustrative only.  (a) Model 1: Distance 35 pc, source diameter 6\arcmin, aspect ratio 1 and $T_e$ $=$ 8,000 K.   (b) Model 2: Distance 35 pc, source diameter 1\arcmin, A $=$ 0.2 and $T_e$ $=$ 50 K.  (c) Model 3: Distance 100 pc, source diameter 6\arcmin, A $=$ 1 and $T_e$ $=$8,000 K.  (d) Model 4: Distance 100 pc, source diameter 1\arcmin, A $=$ 0.2 and $T_e$ $=$ 50 K.  An Aspect Ratio of $<$1 implies a flattened structure along the line-of-sight.  In all four plots the horizontal arrows at the center show the range of electron column density that would produce the observed amplitude of 0.16 mK and the vertical arrow shows the limits of the brightness temperature expected at 23 \&\ 94 GHz at the best fit electron density.  The striking fact that emerges from these examples is that the signals produced by free-free emission from unresolved electron structure in the beams at these two frequencies are closely similar and that they are relatively insensitive to the electron temperature.  Also, the spectrum would appear nearly flat over the frequency range 23 to 94 GHz.  The electron column densities required to produce the HFCE are of order 20\% of the associated HI features for the MI and MII data.

Table 6 summarizes some of the specific parameter values related to these models.  The chosen values entered into Eqtn. 12 are listed in the first four rows and the value for $N_e$ the electron column density in the source structure required to produce the desired HFCE amplitude of 0.16 mK at 23 GHz is shown in the fifth row.  The next four rows list the derived parameters for each model that are associated with the 0.16 mK signals at 23 GHz.  The derived electron volume densities in the source for these models range from 170 to 1,400 cm$^{-3}$.  These imply emission measures of order 1.6 to 12 x 10$^{3}$ cm$^{-6}$ pc.  This may be compared to typical values derived from the Wisconsin H-alpha Mapper (WHAM: Haffner et al. 2003).  Allowing for beam dilution within the WHAM beam of 1\arcdeg\, and converting to Rayleighs, this would produce an observed emission measure of order 20 R, far greater than anything observed at high galactic latitudes, where the typical H${\alpha}$ levels are in the range 0.2 to 0.4 R.  Thus if the {\it WMAP} structure is produced by free-free emission from unresolved electron density enhancements at the edges of HI features the electron temperatures cannot be of order 8,000 K since the corresponding H${\alpha}$ emission is not observed.  Instead we should consider features where the electron temperatures may be as cold as the the HI itself, perhaps as low as 50K, in which case no H${\alpha}$ emission is expected. In that case some mechanism must be invoked to keep the electrons separated from the positive ions to prevent rapid recombination  In the discussion section below we will consider what observations would be needed to test this hypothesis.  The final two rows illustrate the range of electron column densities that are required to produce a signal of 0.16 mK at 23 and 94 GHz.  The values are very close, as can be seen in Fig. 15.  

\section{On the nature of the possible small-scale structures}
A great deal of data relating to the small-scale structure of interstellar structure exist in the literature.  For example, aperture synthesis observations of neutral hydrogen emission and absorption structure (e.g., Braun \& Kanekar 2005; Diamond et al. 1989) reveal very small angular diameter structure to be common.  Also, optical observations of absorption profiles produced by NaI in front of a distant stars or globular clusters (e.g., Meyer \& Lauroesch 1999) show that the absorbing interstellar medium has structure on scales less than ten thousand AU across ($<$0.05 pc).  The average angular diameter of the HI features found in the above study appears to be of order 1\arcdeg\ but it is known from high resolution studies of high-velocity HI, for example by Schwarz \& Oort (1981), Wakker (1991) and Wakker \& Schwarz (1991), that interstellar structures can have angular dimensions as small as 1\arcmin\ in extent within larger complexes.  A comprehensive discussion with many literature references for such "micro-structure" in the diffuse interstellar medium has been presented by Hartquist, Falle \& Williams (2003) and in \S12 we will return to relate their discussion to our model.

In \S10.3 it is hypothesized that unresolved concentrations of electrons in interstellar space could give rise to continuum emission at a level observed by {\it WMAP} summarized in Fig 15 and Table 6.  This raises the question as to whether independent evidence exists that might confirm the hypothesis.  The discovery by Fiedler et al. (1987) of extreme scattering events, or ESEs, indicate that small-scale concentration of electrons do exist.  They estimated that electron clouds with densities of order 1000 cm$^{-3}$ and a linear scale of order 0.5 AU lie along the lines-of-sight to several high-latitude quasars.  The precise mechanism by which the ESEÕs are produced remains enigmatic, as discussed by Walker \& Wardle (1998), and from our point-of-view these features appear to be very much smaller than required to produce the necessary continuum emission observed by {\it WMAP}.  

The likely existence of the necessary small-scale structure produced as a result of MHD turbulence in the interstellar medium has been discussed in a series of theoretical papers, for example by Kowal, Lazarian \& Beresnyak (2007). It is not immediately apparent how to relate the properties predicted by such turbulent models with the data discussed here but that could provide a fruitful avenue for future research 

A recent direct measurement of the electron density along a path length at a high galactic latitude is found in Howk, Sembach \& Savage (2003) who derived a column density of between 1.3 \& 7.9 x 10$^{-19}$ cm$^{-3}$ for a 10 kpc path toward the star vZ1128 at {\it b} = 78.\arcdeg7.  In part they used WHAM data.  The values derived in \S10.3 and summarized in Table 6 and Fig. 15 lie within this range.  However, to reconcile these two facts implies that the bulk of the electron column density found by Howk et al. (2003) is confined to very small-diameter, high density clumps along the line-of-sight.     

To add to the circumstantial evidence in favor of deciding that we are dealing with more than just odd coincidences is the fact that the properties of the structures hypothesized to account for the HFCE as summarized in Fig. 15 and Table 6 is that the implied volume densities of electrons in such features (200 to 1,400 cm$^{-3}$) is in the range of the microstructures discussed by Hartquist, Falle \& Williams (2003) said to be of order 100 AU in extent (equivalent to 5 x 10$^{-4}$ pc. However, this is much smaller than the values pertaining to our model estimates, 0.03 pc at 35 pc distant and 0.01 pc at 100 pc.  Since the Hartquist, Falle \& Wiliams (2003) discussion concerns neutral hydrogen structure, the question then becomes whether electron structures with the sizes and densities implied above also exist in the diffuse interstellar medium. Falle \& Hartquist (2002) have argued that in a cold plasma a variety of slow waves can introduce large density perturbations under suitable conditions and in this regard we note that the diffuse interstellar medium is a cold plasma.  The issue then becomes one of recognizing whether the HI microstructures, when they become ionized, will produce regional enhancements of electron density of the same order as is required to produce the HFCE according to Eqtn. 12.  Clearly, similarities between the parameters implied by models such as summarized in Table 6 and the discussions of Hartquist, Falle \& Wiliams (2003) and Falle \& Hartquist (2002) deserve further consideration.

All of these relationships can individually be described as being due to chance coincidences, which would imply that they have no physical significance. Taken together, however, they raise tantalizing questions.  

\section{Discussion}
This paper has presented what appear to be several intriguing coincidences.  
(a) Small-scale structure observed by {\it WMAP} as summarized in the {\it ILC} map appears to be associated with similar structure found in the distribution of interstellar H emission when mapped in a small velocity band of order 3 km/s wide and separated by 2 km/s in center velocity. 
(b) The angular separation between members of associated pairs is usually of order 0.\arcdeg8.  
(c) In several cases examined closely the associations appear even more dramatic when the HI column densities in specific line width families are mapped.
(d) The angular structure of the two types of features are closely similar showing a cross-correlation coefficient of order unity when allowance is made for beam widths and small offsets in angle.  
(e) In the majority of cases where an association has been noted, HI at more than one velocity may be involved.  
(f) Free-free  emission from interstellar electrons clumped in unresolved structures could produce signals of the same intensity as the high-frequency continuum emission observed by {\it WMAP}.  

Given the pervasive existence of complex HI structure in the Galactic foreground and the equally pervasive nature of the small-scale structure observed by {\it WMAP}, it is important to determine whether confusion can be created in the {\it WMAP} data by previously unrecognized foreground signals associated with interstellar structure.  The {\it WMAP} team (e.g., Hinshaw et al. 2007) went to great lengths to remove contributions from well-understood sources of galactic radio frequency radiation, such as are observed from the galactic disk, the spurs of radio continuum emission, and previously mapped interstellar dust.  There was no {\it a priori} reason for them to expect that HFCE could be produced through processes occurring in otherwise ÒnormalÓ and relatively dust-free interstellar space.  At the high galactic latitudes to which we have confined our study, very little if any cirrus dust is associated with the HI structures in question, although evidence for weak dust emission in HI complex M has recently been reported by Peek et al (2009).  

The overall pattern found in the associations found to date is that small angular offsets exist between paired structures. This is not unexpected, given that there is no {\it a priori} reason to expect that the two types of matter (neutral hydrogen and, for example, clouds of enhanced electron density that may give rise to the HFCE observed by {\it WMAP}), will coexist in identical volumes of space.  More likely is the situation in which the HFCE could be generated at the interfaces between interacting HI features, or at the interface between such features moving with respect to surrounding interstellar plasma.  This would produce a form of limb brightening observed along the region of the interface.  Depending on the orientation of the axis describing the relative motions of the interacting gas masses with respect to the line-of-sight, the associations between HI and HFCE structure may appear coincident in position or will more generally appear slightly offset in angle projected on the sky. 

If the reader is willing to consider that the above sections do not merely list a string of odd coincidences, it may also be necessary to recognize the role of magnetic fields in concentrating electrons through a pinch mechanism or to examine how electrons enhancements could be produced in the first place.  For example, the poorly understood process discussed by Peratt \& Verschuur (2000) may be involved in which a plasma phenomenon can trigger ionization when neutral material streams into a plasma.  Also the role of weak dust emission found toward HI complex M by Peek et al. (2009) may need to be considered and such dust, in turn, will have a bearing on any discussion of the cooling and recombination of electrons, no matter how they are produced.  It is also intriguing to recognize that the volume density of electrons implied by the above models is of the order found in planetary nebulae but that the derived diameters found for these models is of order of 1 to 10\% of the diameters of such nebulae. This would imply the existence of a category of structures smaller than the canonical planetary nebulae.

Whether or not unresolved structures in interstellar electron distribution accounts for the high-frequency continuum emission can be tested by using very different primary beam widths at the same frequencies used by {\it WMAP}.  Note that the Boomerang experiments (Jones et al. 2004) observed the small-scale structure at a frequencies of 145, 245 \& 345 GHz with effective beam width from 11.\arcmin5 to 9.\arcmin1.  If there exist structures in the electron distribution of order of an arcminute in extent the Boomerang data might cast additional light on this model.  However, when the Boomerang beam widths and frequencies are entered in Eqtn. 12 the curves in Fig. 15 are merely shifted to slightly higher electron column densities by only a factor of 1.3 to obtain the same average signal of 0.16 mK.  Given signal-to-noise limitations, this may not be sufficient to test the validity of the model suggested here.

Based on the derived values of the average electron density along the line-of-sight, the models involving the source distances of order 35 to 100 pc share a common property.  Using Eqtn. 12 it turns out that more distant features with the same internal total electron column density produce much weaker signals in a given beam width.  For example, for Models 3 \& 4 at distances of 500 or 1,000 pc the expected brightness of the HFCE will be of order 0.03 \& 0.01 mK respectively at both 23 \&\ 94 GHz and would be below the present threshold of detection.  Thus even in the galactic disk foreground sources consisting of patches of enhanced electron density will dominate the structure observed by {\it WMAP}.

A large-scale comparison meant to detect direct point-to-point correlations between {\it ILC} and galactic HI structures found in the narrow velocity bands shows no clear effect (\S 3), but then there is no obvious evidence that a point-to-point relationship is common in the data.  It is true that foreground galactic HI features exhibit angular scales similar to those observed by {\it WMAP} as epitomized in the {\it ILC} map and, of course, it is well-known that the galactic HI emission covers all of the sky. Therefore near coincidences with HI features will be expected for structure in the {\it ILC} map.  Thus the challenge is one again to seek independent evidence that could either corroborate or negate the claims made above.    

In two instances the relationship to excess soft X-ray structure reinforces the argument that in interstellar space there may indeed exist physical processes capable of producing not only some of the HFCE observed by {\it WMAP} but also excess, albeit weak, soft X-ray emission.  In one case, MI, the result of Gaussian analysis showed that it is HI in a specific line width regime that is most closely associated with the {\it ILC} structure. 

It should be stressed that the above discussion of close associations has been confined to structures found in an area of sky where the {\it WMAP} data required very small if any corrections for possible contributions from dust or synchrotron radiation as can be seen in the analysis of Hinshaw et al. (2007), in particular their Fig. 5.  If one is prepared to consider that what has been found above is not due to chance associations, it will clearly be worth while to compare the HI data with the structures seen in the individual, un-smoothed data obtained by {\it WMAP} at its five frequency bands, in particular at 94 GHz, the highest resolution data set available.  

What has been shown above is that any attempts to confirm or negate the significance of the apparent associations between HI and {\it ILC} structures cannot be accomplished by simply comparing the {\it ILC} data with total HI column densities or even the distribution found by integrating over a 10 km/s velocity range as was done in Paper 1, or by Land \& Slosar (2007).  Instead, a search for closely offset associations needs to make use of HI area maps integrated over, say, 3 km/s and examining them at 2 km/s intervals as was done here.  (Note that HI at 100 K  will produce a line width of 2.2 km/s, which sets a practical limit to the resolution required.)  Furthermore, such maps must be studied over the entire velocity range over which HI is found in any given area of sky, a range can be as large as 300 km/s, in order to determine at what velocity the association of a given {\it ILC} peak with an HI structure appears.  After all, if, for example, the HFCE is produced where two HI features interact then the velocity at which an association is found will depend on the physical conditions of the HI in a given direction.  These interactions will not occur at a specific velocity over the whole sky.  

The physics of interstellar structure, in particular of HI, is clearly complex and a full understanding of such structure as it pertains to creating patches of enhanced electron densities may require invoking the role of plasma and/or magneto-hydrodynamical phenomena as was done by Falle \& Hartquist (2002) and Hartquist, Falle \& Williams (2003).  Perhaps this should also consider magnetic reconnection and/or plasma instabilities in field-aligned HI filaments in order to derive a comprehensive description of the processes that underly the production of HFCE in the direction of diffuse interstellar HI features.  It is even possible that a relatively little known plasma phenomenon known as Marklund convection (Marklund 1979), if it is operating in interstellar space, could account for separation of electrons from their parent ions in HI features. A full discussion of the many implications of the above discussions is way beyond the scope of this paper.

If galactic foreground emission is responsible for at least some of the small-scale structure in the high-frequency continuum radiation observed by {\it WMAP} (and by implication {\it COBE}) it is then very interesting and perhaps a remarkable coincidence that the acoustic spectrum said to describe the existence of sound waves in the early universe should be found to describe the HFCE data obtained by {\it WMAP}.  

In \S11 it was concluded that the WHAM survey data appear to rule out the possibility that dense concentrations of electrons usually associated with a warm ionized medium could be present to produce the observed continuum emission observed by {\it WMAP}.  Instead the observational restraints provided by the WHAM survey lead to the conclusion that the free-free emission that gives rise to a continuum signal as derived from Eqtn. 12 involves cold electrons at temperatures below which a significant fraction would be ionized to produce H${\alpha}$  emission. The most obvious test of the claims made in this paper then comes back to observing the HI and HFCE at higher resolutions to determine if apparent associations between HI and HFCE persist.  The immediate step toward such a comparison will be to examine continuum data obtained by the Planck spacecraft and HI structures observed with the 100-meter Byrd Green Bank Telescope (GBT), both of which have beam widths of order 9\arcmin. However, even before then, those who have ready access to the highest resolution raw {\it WMAP} could determine if the observations at 94 GHz contain structure that can be related to HI data obtained with the GBT.

\section{Conclusions}
Apparent associations between small-scale galactic HI features and structure in the {\it WMAP ILC} data have been presented above and in Paper 1.  The cross-correlation between paired features is extremely high while a general point-to-point relationship expected for directly associated features is not found. It appears that the high-frequency continuum structure observed by {\it WMAP}  may be produced by free-free emission from unresolved clumps of interstellar electrons. The typical offsets between the two forms of emission suggest that the continuum radiation originates at the interface between HI features that are either interacting with one another or with surrounding plasma through which they are moving. In general, the line-of-sight will intersect such features at some angle that would favor our observing an offset between the two types of radiation, although direct positional agreement has been noted in several cases.  The possibility that  these phenomena may be due to more than chance coincidence is reinforced by the fact that the amplitude of high-frequency continuum radiation is at a level expected for free-emission from electrons.  

In summary, the data considered so far suggest several alternative explanations for the apparent near associations between small-scale and highly correlated structures in the distribution of HI and the HFCE observed by {\it WMAP}.  (a) The most obvious one is that the association are all due to chance: This includes 108 close associations reported here and in Paper 1, and another 100 or so noted in a cursory examination of limited areas of the southern skies.  However, the "chance association" conclusion is based on the unstated assumption that we know enough about interstellar processes that we can be sure that it is impossible for a previously unrecognized mechanism to produce HFCE in volumes of space where HI features are interacting with their surroundings, or with one another.  (b) Some of the associations are due to chance and some may be real.  This quickly poses the same dilemma as in (a).  As soon as a few examples are shown to be clearly related to galactic phenomena, such as appears to be the case for MI and MII and the examples given in Paper 1, then the challenge is again to determine what physical mechanisms may be responsible for these associations so as to estimate what fraction of the {\it ILC} peaks may still be assigned a non-galactic origin.  (c) The associations are significant.  If one favors this interpretation, then new aspects of interstellar gas dynamics and physics may be revealed and if the above analysis is any indication then a full discussion will require a great deal more work than can be encompassed in the brief summary presented here. 

In order to understand the physical nature of the apparent associations, the properties of the relevant HI features must be determined by making use of higher angular and velocity resolution data.  Thus some of the key associations discussed above should be studied using HI data obtained with the Green Bank Telescope with a 9\arcmin\ beam and compared to structure in the HFCE to be observed by the Planck spacecraft with a similar beam width.

\acknowledgments
I am particularly grateful to Tom Dame for making available his program for extracting data from the LAB data cube for use in a Mac computer in order to produce area maps using a marvelous program called Transform, which was heartlessly abandoned by IDL.  Without these two programs none of the work involved in making area maps presented here would have been possible on a finite budget in a finite lifetime.  I wish to express my thanks to Wayne Landsman for providing me with the {\it ILC} data in rectangular coordinates.  I am also very grateful for valuable discussions with and the continued encouragement from my wife, Joan Schmelz, who helped tide me through times when I believed that the task undertaken here was akin to tilting at windmills.  It may still be.  Ron Maddelena is thanked for suggesting the statistical test summarized in the \S3 and Butler Burton and Mark Reid for their useful suggestions. Finally, an anonymous referee is thanked for extremely valuable suggestions.

{}

\clearpage

\begin{deluxetable}{ccccccccc}
\tablecolumns{9}
\tablewidth{0pc}
\tablecaption{MII AREA ASSOCIATIONS}
\tablehead{
\colhead{No.} & \colhead{{\it l}}   & \colhead{{\it b}}    & \colhead{{\it ILC} Temp.} & \colhead{HI Velocity}    & \colhead{{\it l}}   & \colhead{{\it b}}    & \colhead{HI amplitude}  & \colhead{Angular offset } \\ {} & {(\arcdeg)} & {(\arcdeg)} & {(mK)} & {(km/s)} & {(\arcdeg)} & {(\arcdeg)} & {(K)} & {(\arcdeg)}}
\startdata
(1) & (2) & (3) & (4) & (5) & (6) & (7) & (8) & (9)\\
65	&	180.7	&	55.2	&	0.238	&	-93	&	181.5	&	54.0	&	1.6	&	1.28	\\
66	&	195.2	&	60.4	&	0.206	&	-14	&	193.0	&	60.4	&	6.5	&	1.09	\\
66	&	195.2	&	60.4	&	0.206	&	-61	&	196.0	&	60.0	&	7.4	&	0.56	\\
67	&	197.2	&	59.6	&	0.195	&	none	&	-	&	-	&	- &	-\\
68	&	193.5	&	63.8	&	0.188	&	8	&	194.0	&	63.0	&	9.4	&	0.82	\\
69	&	196.8	&	66.2	&	0.178	&	-29	&	194.6	&	66.0	&	12.7	&	0.91	\\
70	&	192	&	57.5	&	0.165	&	-93	&	192.9	&	58.0	&	1.8	&	0.70	\\
70	&	192	&	57.5	&	0.165	&	-45	&	193.0	&	57.0	&	14.5	&	0.73	\\
71	&	209.4	&	62.6	&	0.16	&	-39	&	208.0	&	62.1	&	23.7	&	0.82	\\
71	&	209.4	&	62.6	&	0.16	&	-35	&	208.0	&	63.1	&	18.1	&	0.82	\\
72	&	196.2	&	62.9	&	0.159	&	-65	&	196.0	&	63.6	&	8.6	&	0.71	\\
73	&	194.2	&	54	&	0.159	&	-97	&	193.0	&	53.0	&	3.4	&	1.26	\\
74	&	206.4	&	55	&	0.156	&	4	&	206.9	&	55.0	&	15.5	&	0.29	\\
75	&	200.9	&	58.6	&	0.151	&	-35	&	201.5	&	59.9	&	10.7	&	1.34	\\
76	&	187.4	&	63.8	&	0.148	&	-97	&	186.1	&	64.0	&	4.0	&	0.63	\\
76	&	187.4	&	63.8	&	0.148	&	-12	&	188.0	&	64.0	&	5.0	&	0.33	\\
77	&	190.5	&	56.3	&	0.144	&	-12	&	189.6	&	55.6	&	3.3	&	0.86	\\
78	&	193.7	&	56.2	&	0.142	&	-22	&	193.0	&	55.6	&	6.5	&	0.72	\\
79	&	201.2	&	56.6	&	0.139	&	-49	&	201.5	&	56.5	&	33.1	&	0.19	\\
80	&	202.4	&	65.2	&	0.137	&	-57	&	203.6	&	65.6	&	7.3	&	0.64	\\
80	&	202.4	&	65.2	&	0.137	&	-36	&	202.0	&	65.1	&	8.6	&	0.20	\\
81	&	204.5	&	67.4	&	0.126	&	-29	&	205.7	&	66.5	&	16.5	&	1.01	\\
82	&	187.7	&	55	&	0.125	&	-105	&	187.2	&	55.5	&	1.7	&	0.58	\\
83	&	195.85	&	55.5	&	0.124	&	-12	&	196.6	&	55.5	&	5.3	&	0.42	\\
84	&	208.4	&	58.5	&	0.122	&	none 	&	- &	- 	&	-	&	-\\
85	&	199	        &	53.4	&	0.116	&	-6	&	198.6	&	53.4	&	25.4	&	0.24	\\
86	&	199.9	&	50.5	&	0.116	&	-93	&	198.0	&	51.0	&	2.6	&	1.31	\\
87	&	186.15	&	50.4	       &	0.113	&	-18	&	184.9	&	50.5	&	7.0	&	0.80	\\
88	&	186.75	&	51.45	&	0.106	&	-35	&	186.5	&	51.5	&	4.9	&	0.16	\\
89	&	182.15	&	51.35	&	0.094	&	12	&	183.5	&	50.5	&	7.2	&	1.18	\\
90	&	189.5	&	52.1  	&	0.093	&	-43	&	189.5	&	51.7	&	2.9	&	0.40	\\
91	&	183.15	&	66.8	       &	0.085	&	-65	&	183.6	&	65.5	&	8.2	&	1.31	\\
91	&	183.15	&	66.8   	&	0.085	&	-6	&	183.8	&	67.5	&	5.5	&	0.75	\\
92	&	194.15	&	50.25	&	0.085	&	-18	&	195.6	&	50.5	&	7.0	&	0.96	\\
93	&	181.85	&	64.85	&	0.073	&	-6	&	183.0	&	65.0	&	5.5	&	0.48	\\
94	&	189.4	&	60.45	&	0.071	&	-65	&	189.6	&	60.1	&	5.6	&	0.36	\\
94	&	189.4	&	60.45	&	0.071	&	-57	&	189.0	&	60.6	&	10.6	&	0.25	\\
94	&	189.4	&	60.45	&	0.071	&	-6	&	189.0	&	60.6	&	6.5	&	0.22	\\
95	&	196.55	&	52	&	0.069	&	-101	&	197.0	&	52.0	&	1.6	&	0.25	\\
95	&	196.55	&	52	&	0.069	&	-77	&	195.4	&	52.0	&	3.4	&	0.74	\\
95	&	196.55	&	52	&	0.069	&	-18	&	196.5	&	51.4	&	8.4	&	0.60	\\
95	&	196.55	&	52	&	0.069	&	-6	&	196.5	&	52.5	&	7.8	&	0.45	\\
96	&	191.35	&	54.4	&	0.065	&	none	&	-	&	-	& -	& -	\\
97	&	205.85	&	57.9	&	0.064	&	0	&	205.0	&	57.9	&	6.2	&	0.45	\\
98	&	190.7	&	66.8	&	0.063	&	-29	&	191.4	&	67.1	&	33.8	&	0.41	\\
98	&	190.7	&	66.8	&	0.063	&	-14	&	189.1	&	66.0	&	5.4	&	1.02	\\
\enddata
\end{deluxetable}

\clearpage

\begin{deluxetable}{ccccccccc}
\tablecolumns{9}
\tablewidth{0pc}
\tablecaption{Top Ten {\it ILC} Peaks}
\tablehead{
\colhead{No.} & \colhead{{\it l}}   & \colhead{{\it b}}    & \colhead{{\it ILC} Temp.} & \colhead{HI Velocity}    & \colhead{{\it l}}   & \colhead{{\it b}}    & \colhead{HI amplitude}  & \colhead{Angular offset } \\ {} & {(\arcdeg)} & {(\arcdeg)} & {(mK)} & {(km/s)} & {(\arcdeg)} & {(\arcdeg)} & {(K)} & {(\arcdeg)}}
\startdata
(1) & (2) & (3) & (4) & (5) & (6) & (7) & (8) & (9)\\
99	&	171.35	&	-45.7	&	0.306	&	-18	&	171	&	-45.9	&	28.6	&	0.32	\\
99	&	171.35	&	-45.7	&	0.306	&	-30	&	170.9	&	-45.5	&	4.2	&	0.37	\\
100	&	184.1	&	-54.5	&	0.259	&	-50	&	181.55	&	-55	&	0.7	&	1.56	\\
100	&	184.1	&	-54.5	&	0.259	&	20	&	185	&	-56	&	2.8	&	1.59	\\
101	&	160.2	&	-58.25	&	0.261	&	-58	&	160	&	-58.45	&	1.2	&	0.23	\\
101	&	160.2	&	-58.25	&	0.261	&	-12	&	161.5	&	-58	&	21.9	&	0.73	\\
102	&	176.6	&	-48.95	&	0.259	&	-8	&	177.1	&	-49.45	&	51.3	&	0.60	\\
102	&	176.6	&	-48.95	&	0.259	&	12	&	177.95	&	-49.4	&	63.5	&	0.99	\\
103	&	164.75	&	-57.45	&	0.252	&	-6	&	164.1	&	-56.5	&	34.6	&	1.01	\\
103	&	164.75	&	-57.45	&	0.252	&	-66	&	164	&	-57	&	1.3	&	0.60	\\
104	&	93.45	&	-37.2	&	0.251	&	-8	&	93	&	-38	&	65.2	&	0.88	\\
104	&	93.45	&	-37.2	&	0.251	&	8	&	94.5	&	-36.9	&	8.1	&	0.89	\\
104	&	93.45	&	-37.2	&	0.251	&	14	&	92.95	&	-37.5	&	4.1	&	0.50	\\
104	&	93.45	&	-37.2	&	0.251	&	24	&	93.5	&	-36.9	&	2.6	&	0.30	\\
105	&	94.3	&	-40.8	&	0.249	&	18	&	94.05	&	-40	&	2.4	&	0.82	\\
105	&	94.3	&	-40.8	&	0.249	&	-14	&	92.45	&	-40.9	&	22.6	&	1.40	\\
106	&	179.95	&	-54.65	&	0.248	&	none	&	- &	-	& -	&	- \\
107	&	37.8	&	42.15	&	0.246	&	-38	&	37.05	&	42	&	2.2	&	0.58	\\
107	&	37.8	&	42.15	&	0.246	&	-10	&	37.9	&	41.5	&	12.6	&	0.65	\\
108	&	189.55	&	-51.5	&	0.244	&	-6	&	188	&	-52.1	&	58.6	&	1.14	\\
108	&	189.55	&	-51.5	&	0.244	&	10	&	189	&	-51	&	55.8	&	0.61	\\

\enddata
\end{deluxetable}

\clearpage

\begin{deluxetable}{cccccccc}
\tablecolumns{8}
\tablewidth{0pc}
\tablecaption{Average Separations between Paired {\it ILC}-HI Peaks}
\tablehead{
\colhead{Ref.} & \colhead{{\it l} range}   & \colhead{{\it b} range}    & \colhead{{\it ILC} peaks (mK)} & \colhead{No.}    & \colhead{Pairs}   & \colhead{HI peaks}    & \colhead{Separation (\arcdeg)}  }
\startdata
1	&	60\arcdeg - 180\arcdeg	&	30\arcdeg - 70\arcdeg	&	$>$0.100 	&	64	&	62	&	83	&	0.93 $\pm$ 0.55\\
2	&	180\arcdeg - 210\arcdeg	&	50\arcdeg - 70\arcdeg	&	$>$0.063	&	34	&	31	&	48	&	0.67 $\pm$ 0.35\\
3	&	All longitudes	&	$|$b$|$ $=$ 30\arcdeg - 70\arcdeg	&	$>$0.244	&	10	&	9	&	21	&	0.79 $\pm$ 0.40\\
\enddata
\end{deluxetable}

\clearpage

\begin{deluxetable}{ccccc}
\tablecolumns{5}
\tablewidth{0pc}
\tablecaption{Average Properties of Gaussians for MII \& MI}
\tablehead{
\colhead{Source name} & \colhead{Linewidth} & \colhead{Peak Temperature} & \colhead{Center velocity} & \colhead{Column density} \\ {} & {(km/s)} & {(K)} & {(km/s)} & {($10^{18}  cm^{-2}$)} 
}   
\startdata
(1) & (2) & (3) & (4) & (5)\\
MI	&	34	& 0.42 $\pm$ 0.40	& -106.5 $\pm$ 8.6	& 26.3 $\pm$ 24.8\\
MI	&	21.1 $\pm$ 1.6 &	0.76 $\pm$ 0.71 & -114.2 $\pm$ 4.7 & 29.0 $\pm$ 27.1\\
Other MI	& 13.8 $\pm$ 2.2 & 0.99 $\pm$ 1.00 & -117.2 $\pm$ 7.2 & 25.8 $\pm$ 26.7 \\
components	& 6.3 $\pm$ 2.4 & 0.54 $\pm$ 0.51 & -117.1 $\pm$ 5.5	& 7.0 $\pm$ 8.8\\
MII	&	34	&  0.32 $\pm$ 0.22 &	-81.6 $\pm$ 4.1 & 19.6 $\pm$ 13.5	\\
MII	&	20.2 $\pm$ 3.4 & 1.00 $\pm$ 0.79	& -80.8 $\pm$ 5.5 & 35.7 $\pm$ 27.2\\
\enddata
\end{deluxetable}

\clearpage

\begin{deluxetable}{cccc}
\tablecolumns{4}
\tablewidth{0pc}
\tablecaption{Correlation Coefficients}
\tablehead{
\colhead{Source name} & \colhead{HI velocity} & \colhead{R in longitude} & \colhead{R in latitude}     }
\startdata
Overlap	&	-87 km/s	&	0.81 unshifted	&	0.89	\\
(Paper 1)	&		&	0.94 offset by 0.\arcdeg7	&		\\
	&	-118 km/s	&	0.94	&	0.87	\\
	&		&		&	0.94 offset by 0.\arcdeg7	\\
	&		&		&		\\
Source \#74	&	+1 to +3 km/s	&	0.71 unshifted	&	0.86	\\
(Fig. 14c)	&		&	0.92 offset by 0.\arcdeg5	&		\\
	&		&		&		\\
Source \#79	&	-50 to Ð48 km/s	&	0.87 unshifted	&	0.87	\\
(Fig 14a)	&		&	0.94 offset by 0.\arcdeg5	&		\\
	&		&		&		\\
In between	&	-40 to Ð38 km/s	&	0.86	&	0.72	\\
(Fig. 14b)	&		&		&		\\
&		&		&		\\
Abutting peaks  	&	-102 km/s	&	0.95 offset by 	&		\\
(Fig. 13)	&		&	1.\arcdeg1 at an angle	&		\\
\\MI (Fig. 5b)	&	-110 km/s	&	0.84	&	0.98 resolved	\\
22 km/s wide component	&		&	0.89 offset 1\arcdeg	&	offset 1\arcdeg in long	\\
 &		&		&		\\
MI 	&	-10 to Ð5 km/s	&	0.97 resolved	&	0.94 	\\
(Fig. 7b)	&		&	offset 0.\arcdeg8 in lat	&	offset 0.\arcdeg8 in lat	\\
&		&		&		\\
\enddata

\tablecomments{Due to beam width differences, R$=$0.93 for unresolved features is equivalent to R$=$1.00, see text.  Structures are unresolved unless otherwise noted}

\end{deluxetable}

\begin{deluxetable}{ccccc}
\tablecolumns{5}
\tablewidth{0pc}
\tablecaption{Model Parameters}
\tablehead{
\colhead{Parameter} & \colhead{Model 1} & \colhead{Model 2} & \colhead{Model 3}  & \colhead{Model 4}   }
\startdata
Chosen parameters:	&		&		&	&	\\
Electron temperature (K)	&	8000	&	50	&	8000	  & 50\\
Angular width (\arcmin)	&	6	&	1	&	6	& 1 \\
Aspect Ratio	&	1	&	1	&	0.2	&   0.2\\
Distance (pc)	&	35	&	35	&	100	& 100\\
 &		&		&	&	\\
For  $N_e$ (10$^{18}$ cm$^{-2}$) the	&	34	&	34	&	26   & 25	\\
derived parameters are:	&		&		&	&	\\
$T_B$ at 23 GHz (mK)	&	0.16	&	0.16	&	0.16 &     0.16	\\
$T_B$ at 94 GHz mK	&	0.13	&	0.11	&	0.13  &     0.10	\\
Volume density (cm$^{-3}$) & 184  &  1,100 & 250 & 1,400\\
Average $n_e$ in l.o.s (cm$^{-3}$)	&	0.32	&	0.32	&	0.09	 &      0.08\\
For $T_B$($\nu$) $=$ 0.16 mK,	&		&		&	&	\\
the required $N_e$ values are: & & & & \\
at 23 GHz (10$^{18}$ cm$^{-2}$)	&	34	&	34	&	26	&        25\\
at 94 GHz (10$^{18}$ cm$^{-2}$)	&	37	&	42	&	28	&        31\\
\enddata
\end{deluxetable}

\clearpage
\begin{figure}
\figurenum{1}
\epsscale{.8}
\plotone{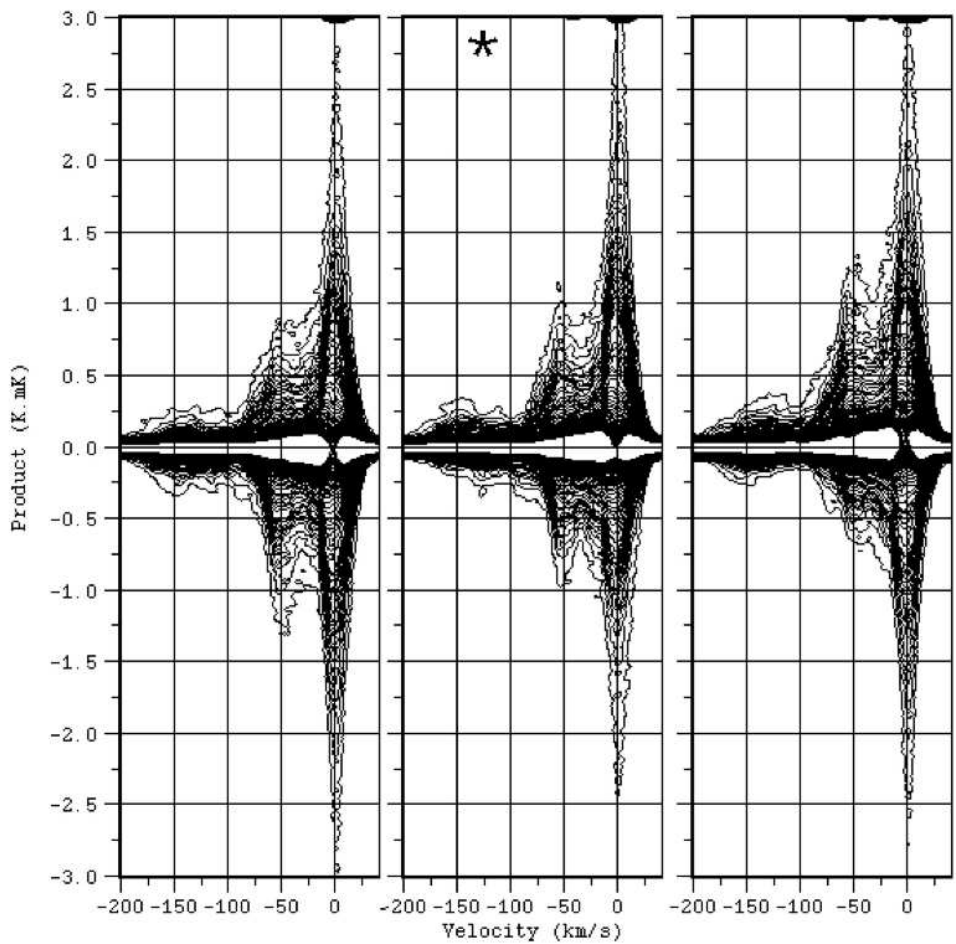}
\caption {Examples of the statistical tests for the relationship between the HI and {\it ILC} data for the northern hemisphere Target Area.  The products derived from Eqtns. (4), (3) \& (2) are shown from left to right respectively.  The product values were initially calculated in intervals of 0.01 K.mK and smoothed to allow this contour map representations.  Contour intervals are 2:20@2, 25:50@5, 60:300@10 K.mK.  The peak values inside the blank areas reach as much as 9000 K.mK.  The asterisk in the center plot marks the case where the HI data are compared with the {\it ILC} data for the same region.  See text for discussion of these plots.}
\end{figure}

\clearpage

\begin{figure}
\figurenum{2}
\epsscale{.8}
\plotone{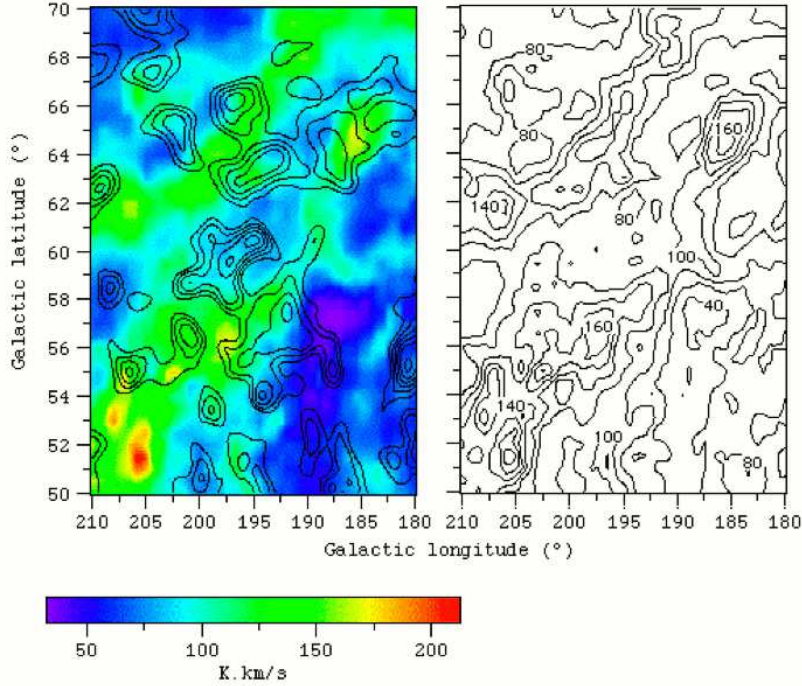}
\caption{ (a) The left-hand figure shows the total HI content integrated from $-$150 to $+$30 km/s for an area encompassing HI feature MII seen at ({\it l,b})$=$ (184\arcdeg,65\arcdeg). The {\it ILC} contour levels overlain are from +0.03 mK in steps of 0.02 mK.  In order to illustrate the challenges posed by comparing {\it ILC} data to the HI data, the right-hand figure (b) displays the HI total column density data in contour map form.  In the identification of HI peaks associated with specific {\it ILC} peaks, the HI data in contour maps were used and the positions and amplitudes determined from that database.  This figure does reveal several close associations in the area around ({\it l,b}) $=$ (204\arcdeg,55\arcdeg), which are shown in detail in Fig. 3.}
\end{figure}

\clearpage

\begin{figure}
\figurenum{3}
\epsscale{.7}
\plotone{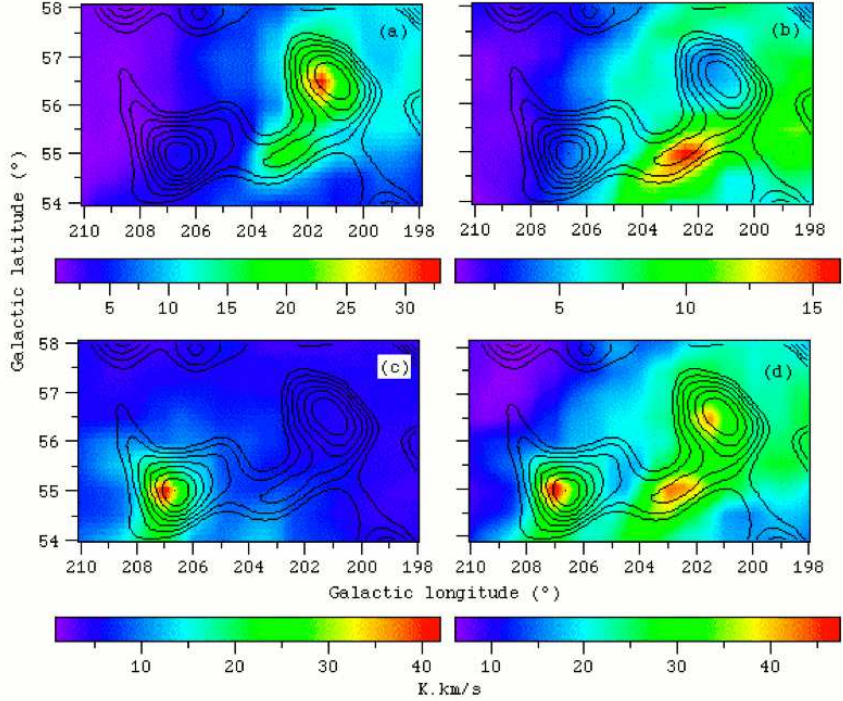}
\caption {This illustrates how the {\it ILC} contours around ({\it l,b}) = (204\arcdeg,55\arcdeg) relate to the existence of three distinct peaks in the HI emission derived from data at three different velocities that are located in the area illustrated in Fig. 2.  (a) The HI data in the velocity range $-$50 to $-$48 km/s are compared to the {\it ILC} contours.  The HI structure at ({\it l,b}) = (201\arcdeg,56.\arcdeg6) is hardly visible at all in the map of total HI content in Fig. 2(a) but is directly associated with the {\it ILC} peak, catalogued as Source \#79 in Table 1.  (b) The HI data in the velocity range $-$40 to $-$38 km/s are compared to the {\it ILC} contours and reveals a clear association between HI and {\it ILC} structure at ({\it l,b}) = 202\arcdeg,54.\arcdeg6.   (c) The HI data are integrated from $+$1 to $+$3 km/s and the contours in the two forms of emission are virtually perfectly aligned.  This is source \#74 in Table 1.  (d) The data in the first three plots are here combined into one plot, which shows the very high degree of association between the two forms of emission for what is clearly a complex morphology at three distinct velocity regimes.}
\end{figure}

\clearpage

\begin{figure}
\figurenum{4}
\epsscale{1.2}
\plottwo{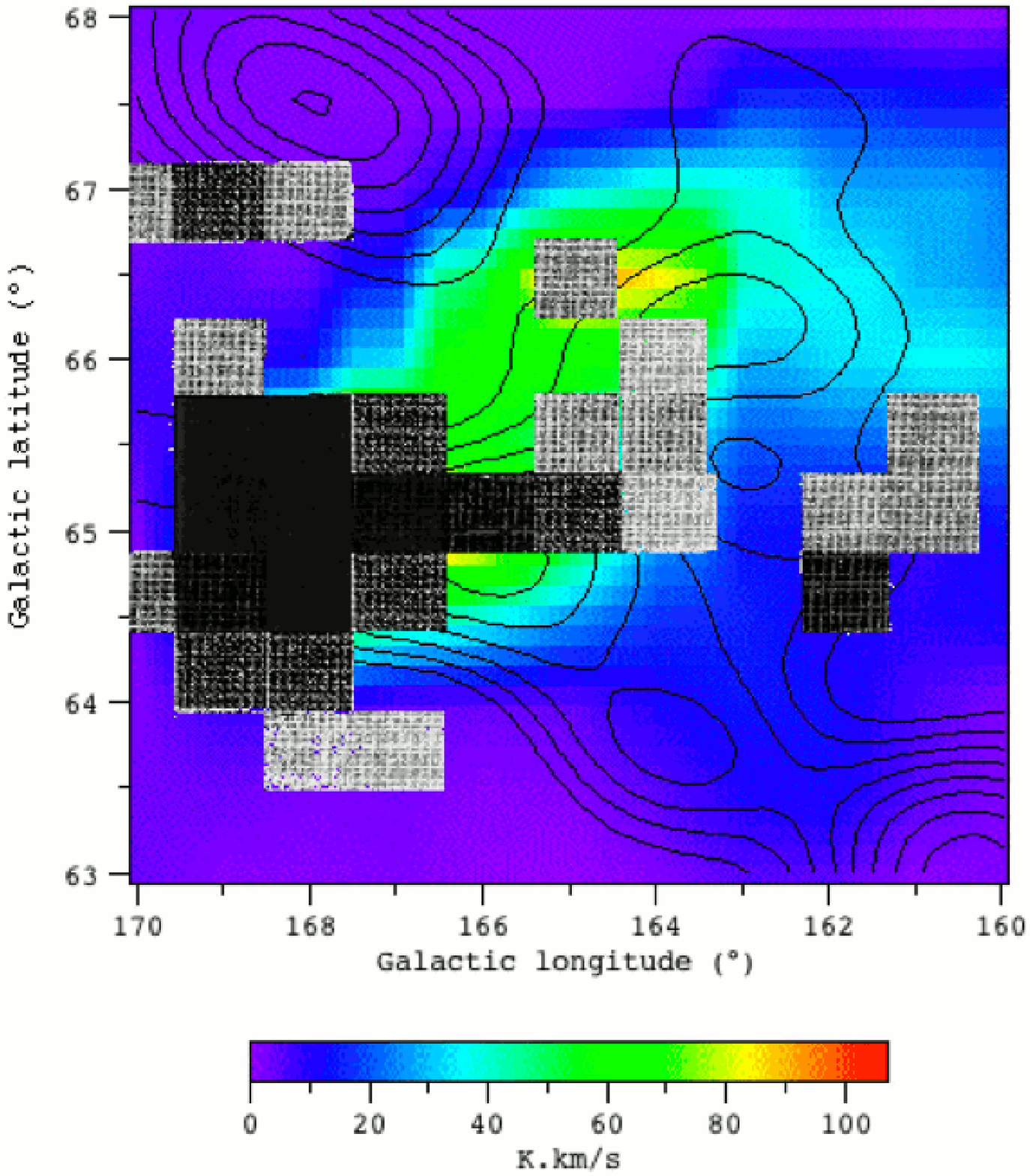}{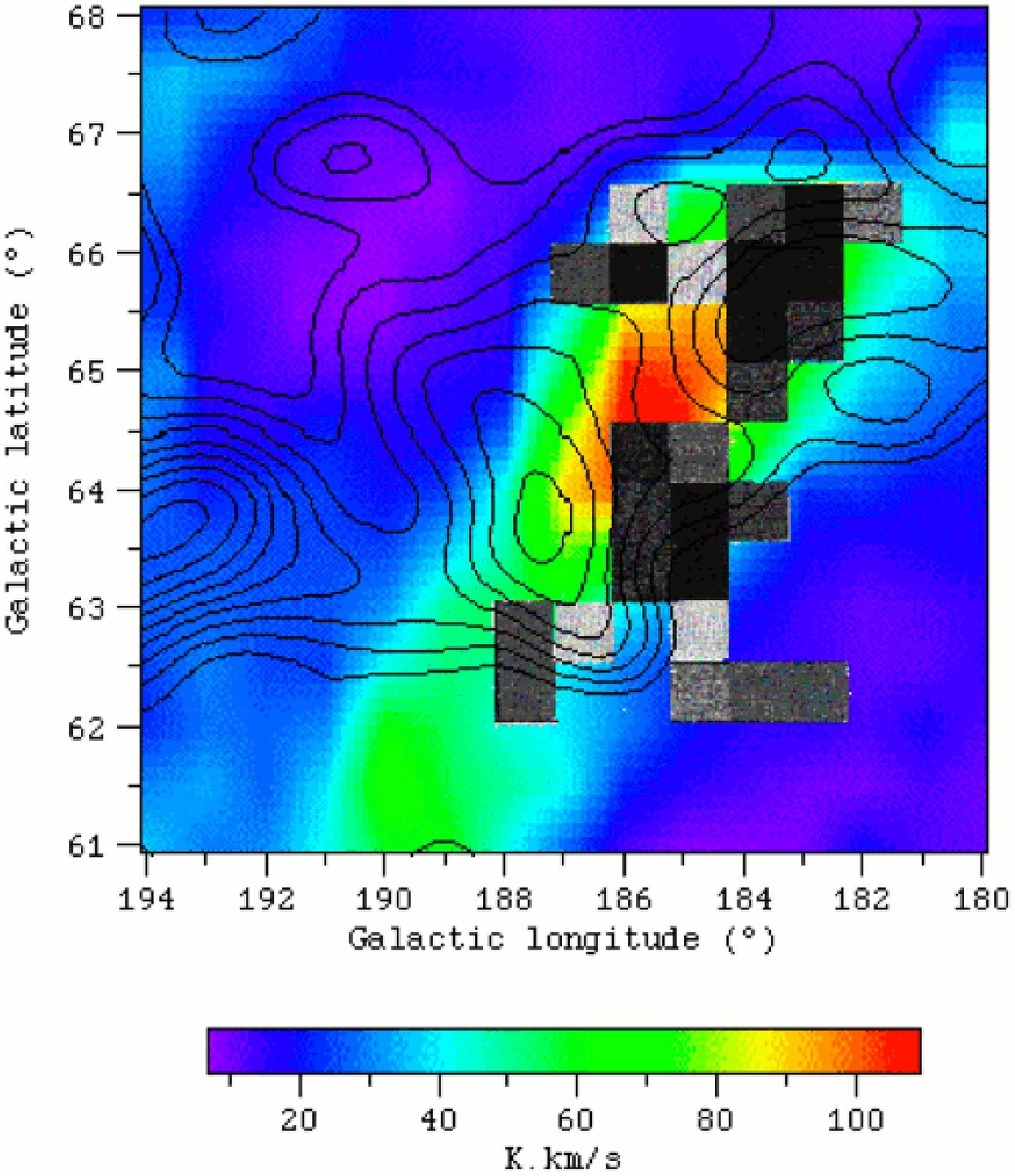}
\caption {(a) The left-hand figure displays the total HI column density for MI from $-$140 to $-$80 km/swith shaded pixels indicating the presence of excess soft X-ray emission at 1/4 keV derived from the data of Herbstmeier et al. (1995).  (b) The right-hand figure displays the total HI column density for MII integrated from $-$120 to $-$55 km/s, which covers the great majority of HI emission at any velocity in this area of sky with the {\it ILC} contours from $+$0.02 mK in intervals of 0.02 mK overlain.  The shaded pixels again indicate the excess soft X-ray emission derived from Fig. 7b of Herbstmeier et al. (1995).  Note that the peaks found in all three forms of emission are slightly offset from one another in contrast to the case in (a) where they are closely aligned.
}
\end{figure}

\clearpage

\begin{figure}
\figurenum{5}
\epsscale{.7}
\plotone{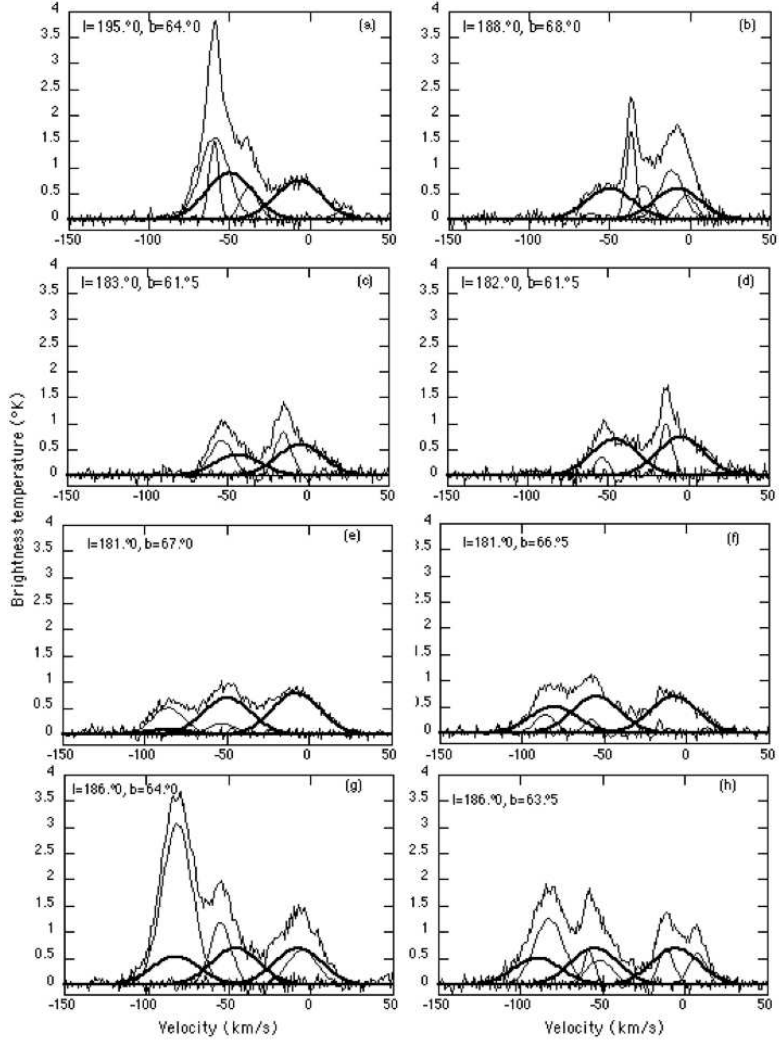}
\caption {Eight frames showing typical profiles in the area toward MII with the Gaussians fit to those profiles.  In the top four frames no emission associated with MII is found.  In the lower four frames emission from MII can be seen around $-$80 km/s.  The Gaussian components indicated by thick lines correspond to the underlying features with a linewidth of 34 km/s.  At low velocities these produce a near perfect fit to many of the observed emission profiles.  In frame (g) the profile at ({\it l,b}) $=$ (186\arcdeg, 64\arcdeg) shows the manner in which the 21 km/s wide component dominates the emission from MII. }
\end{figure}

\clearpage

\begin{figure}
\figurenum{6}
\epsscale{0.7}
\plotone{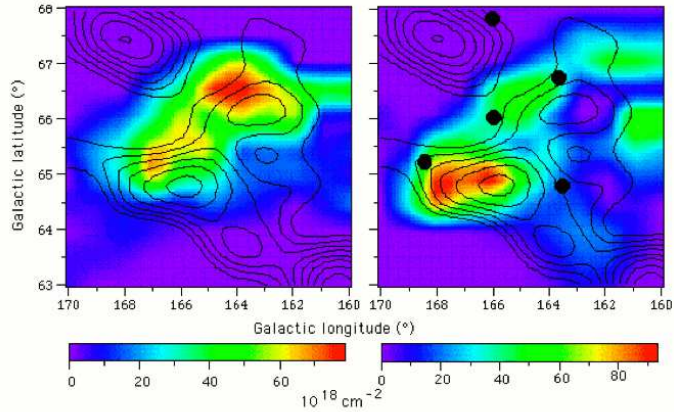}
\caption {The left-hand figure, (a) displays the HI column density for the 34 km/s wide HI component in the direction of HI feature MI compared to the {\it ILC} contours from +0.06 mK in intervals of 0.02 mK.  The double structure in the HI is uniformly offset from the double feature in the {\it ILC} data.  In Paper 1 the total column density of the HI emission was compared with the {\it ILC} structure but in this plot it is evident that the physical properties of the HI associated with each of the two {\it ILC} peaks is distinctly different.   This conclusion is reinforced in the right-hand plot, (b), showing the HI column density for components in the 21 km/s regime compared to the {\it ILC} contours.  This component, indicative of HI at a temperature of about 8,000 K, is very closely associated with the main {\it ILC} peak.  Also shown as filled circles are direction in which Tufte et al. (1998) detected H${\alpha}$ emission. Note that the peaks in all three categories are offset from one another, see text.}
\end{figure}

\clearpage

\begin{figure}
\figurenum{7}
\epsscale{.9}
\plotone{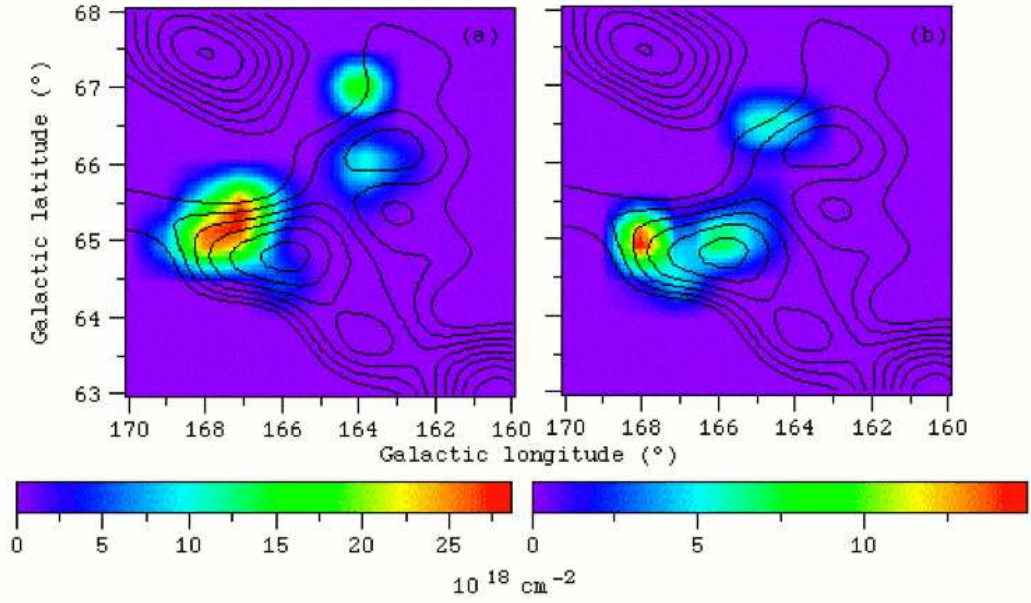}
\caption {Two families of narrow linewidth components are also found toward MI and they are here plotted with respect to the {\it ILC} contours which are the same as for Fig. 6. (a) Column density map for components with linewidths between 9 \& 15 km/s.  (b) Column density map for narrow components with linewidth $<$9 km/s.  Both these families appear to be related to the presence of the brighter of the two {\it ILC} peaks associated with MI. }
\end{figure}

\clearpage

\begin{figure}
\figurenum{8}
\epsscale{.8}
\plotone{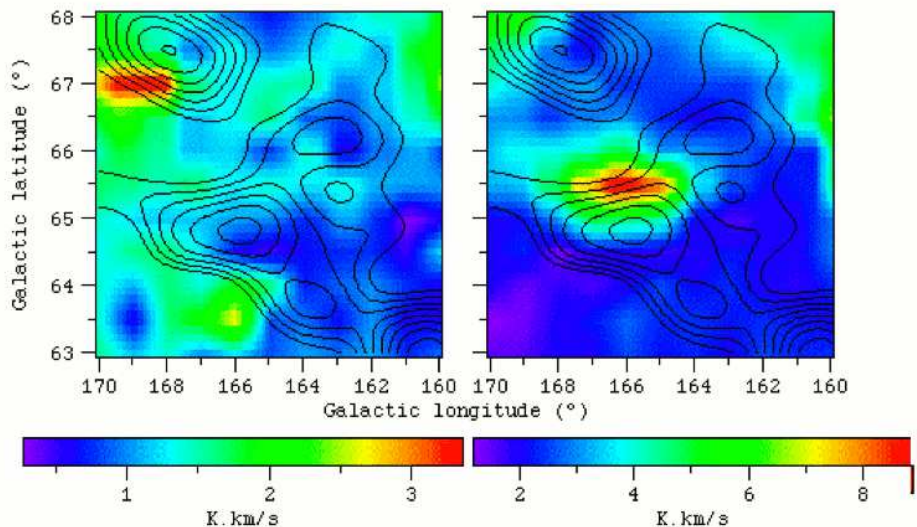}
\caption {Two ({\it l,b}) maps for the MI area again shown with respect to the same {\it ILC} contours used in the previous two figures.  (a) In the left-hand figure a bright HI feature found by integrating between $+$5 and $+$10 km/s appears to be precisely aligned with the small patch of excess soft X-ray emission seen in Fig. 4a.  See text.  (b) The right-hand figure shows an HI feature found by integrating between $-$10 and $-$5 km/s with virtually the same angular extent as brightest {\it ILC} peak in this area is located just to the north of that peak by about 0.\arcdeg5.  This pattern of finding HI features at two distinct velocities associated with a given {\it ILC} peak was found to be common, and this example is particularly striking.  See text. }
\end{figure}

\clearpage

\begin{figure}
\figurenum{9}
\epsscale{0.7}
\plotone{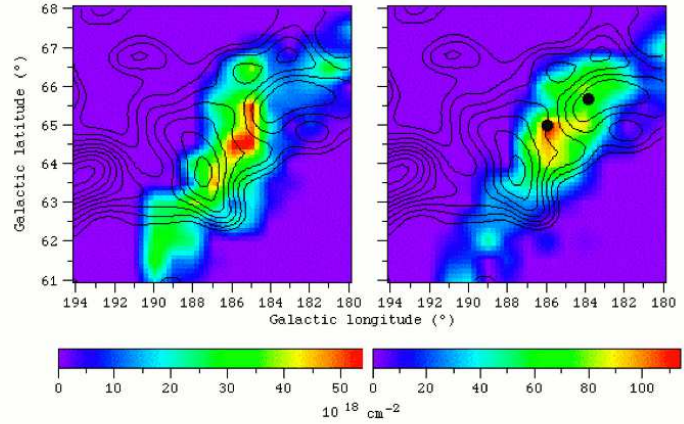}
\caption {(a) At the left, the morphology of the column density of the 34 km/s wide HI component associated with MII compared to the {\it ILC} contours.  (b) At the right, the morphology of the column density of a family of linewidths in the 20 km/s range compared to the {\it ILC} contours from $+$0.02 mK in intervals of 0.02 mK overlain. Also shown as filled circles are direction in which Tufte et al. (1998) detected H${\alpha}$ emission.  Unlike the case for the data in Fig. 6, here one of the H${\alpha}$ detection is closely aligned to the HI.}
\end{figure}

\clearpage

\begin{figure}
\figurenum{10}
\epsscale{.9}
\plotone{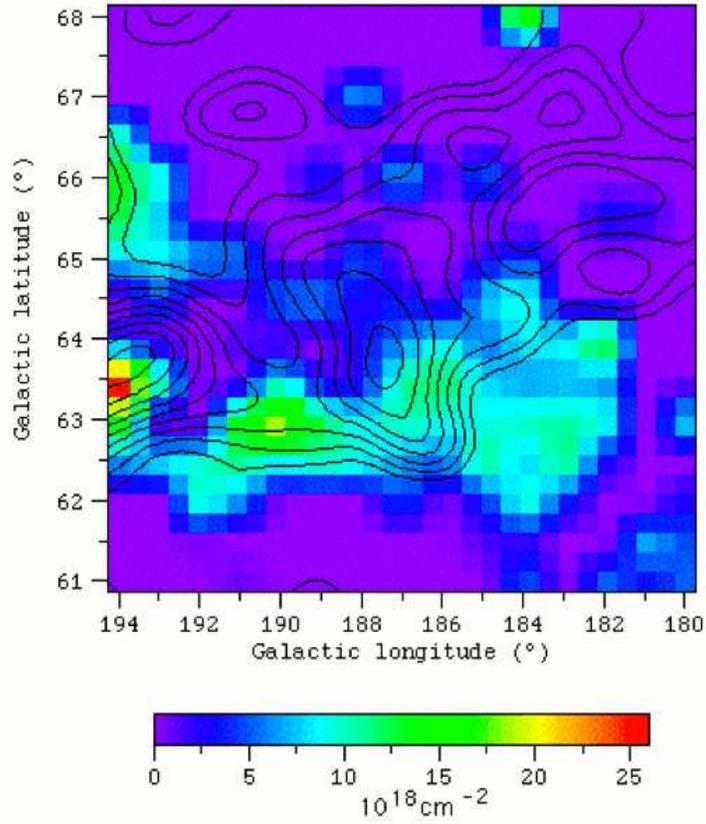}
\caption {Gaussian analysis of the HI emission profiles in the area of MII revealed unexpected features at positive velocities between $+$5 to $+$20 km/s.  Here the HI column density of these components is plotted with the same {\it ILC} contours as in Fig. 9 overlain.  It is striking that the bright peak at ({\it l,b}) $=$ (194\arcdeg,63.\arcdeg4) appears associated with a distinct {\it ILC} peak.  Also a low level of emission at these velocities borders the brightest {\it ILC} peak in the area.}
\end{figure}

\clearpage

\begin{figure}
\figurenum{11}
\epsscale{.8}
\plotone{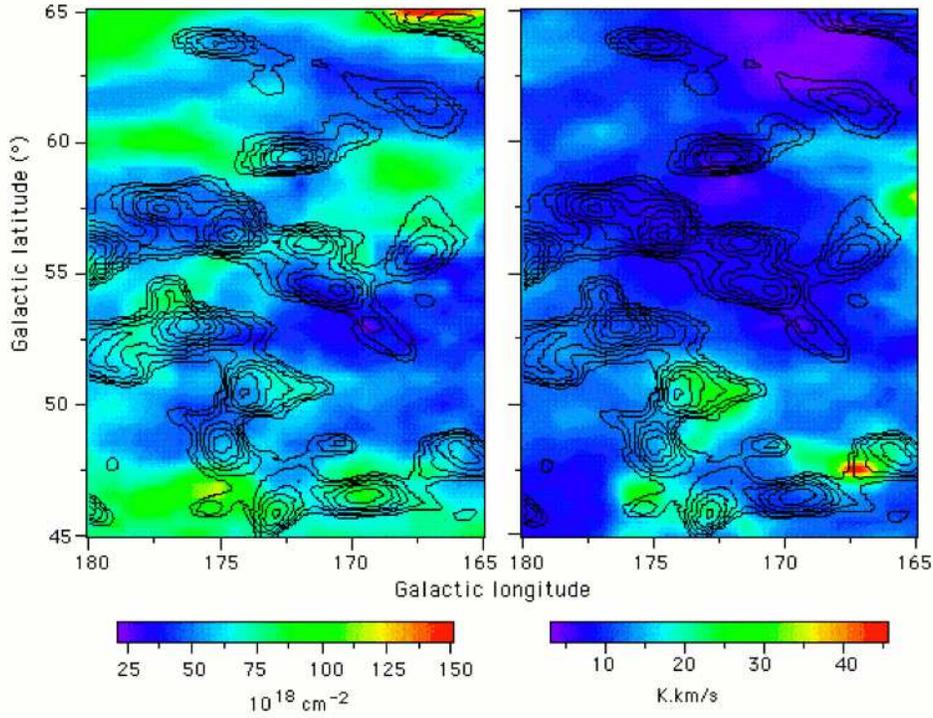}
\caption {(a) The total HI column density for an area just below MI compared to the {\it ILC} contours level used before.  Due to the dynamic range of this plot, few associations stand out.  (b) The same area showing the integrated HI brightness from the -20 to -10 km/s at which resolution the associations begin to manifest.  In Fig. 12 a close-up view of the structure around ({\it l,b}) $=$ (173\arcdeg,50.\arcdeg5) shows a very high degree of association.}
\end{figure}

\clearpage

\begin{figure}
\figurenum{12}
\epsscale{.9}
\plotone{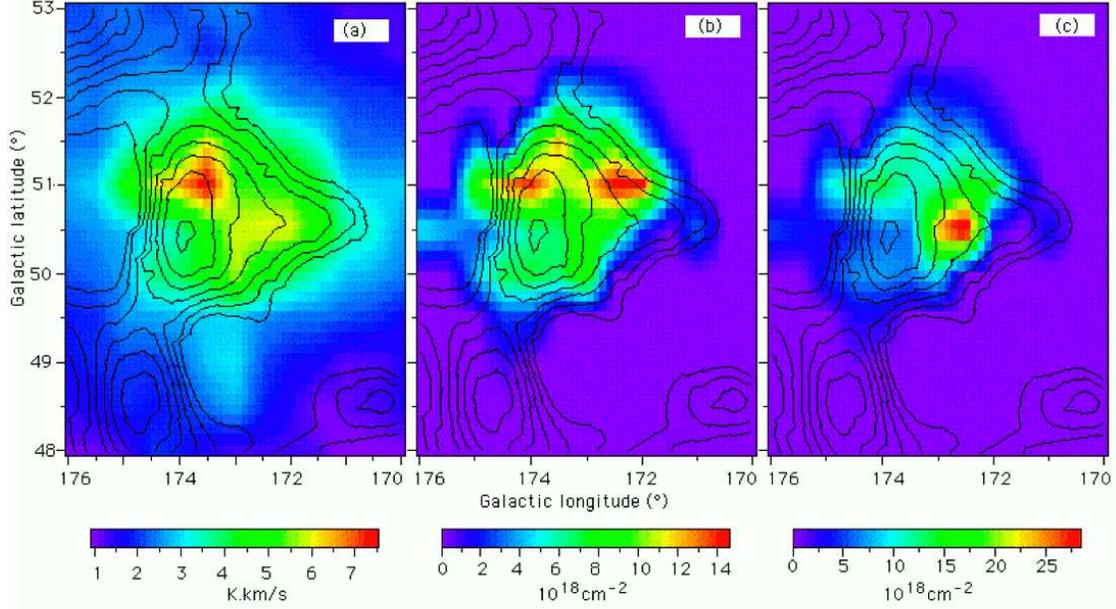}
\caption {Three detailed views of the associated structures around ({\it l,b}) $=$ (173\arcdeg,50.\arcdeg5) seen in Fig. 11.  (a) A detailed view of the association between the HI brightness at $-$19 km/s (integrated over a 2 km/s bandwidth) compared to the same {\it ILC} contours as in previous figures.  This is listed as source \#7 in Table 1 of Paper 1.  The outer boundary of the HI feature corresponds virtually perfectly to the {\it ILC} contours on three sides.  (b) The HI column density found in the Gaussian analysis of a narrow linewidth component corresponding to the emission peak at $-$19 km/s is displayed with the {\it ILC} contours overlain.  The high degree of correlation between the two forms of emission and the fact that they define the same area noted in (a) is even more striking in this plot.  (c) Here both narrow linewidth components found in the direction marked by the bright peak are added together.  In (b) only one of those components was included but in (c) the combined value of column densities for this double Gaussian (which cannot be unambiguously separated in the available HI data) overwhelms the morphology. }
\end{figure}

\clearpage

\begin{figure}
\figurenum{13}
\epsscale{.9}
\plotone{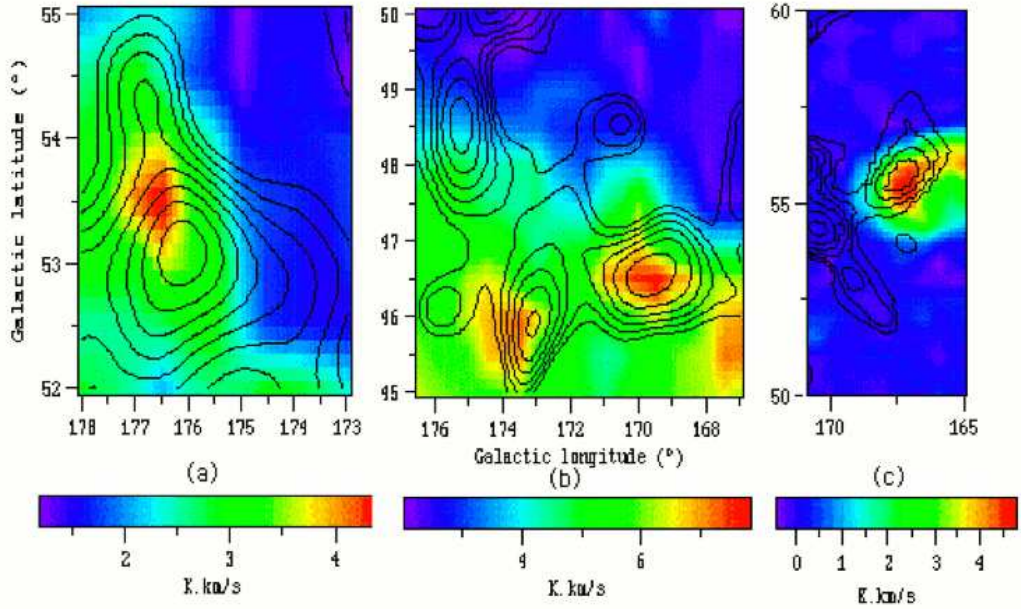}
\caption {Four more examples in the area just south of MI that show clear associations between the {\it ILC} structure (same contour levels as in the previous figures) and HI emission brightness.  Here the two forms of emission are either well aligned or slightly offset by up to 0.\arcdeg5.  The two figure at the left ((a) and (b)) are both channel maps made at $+$2 km/s and the figure at the right (c) shows data at -136 km/s.}
\end{figure}

\clearpage

\begin{figure}
\figurenum{14}
\epsscale{.7}
\plotone{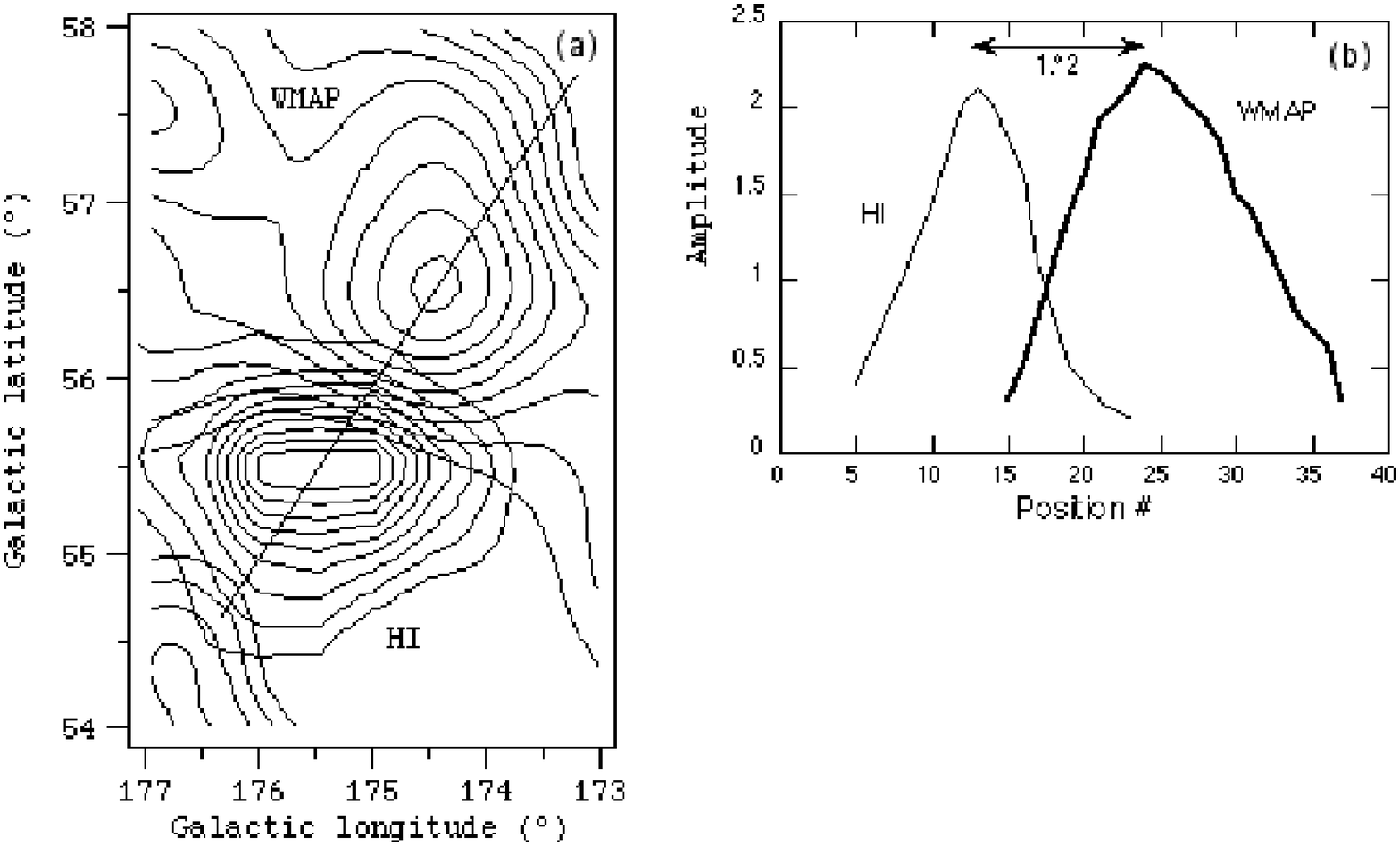}
\caption {An example of closely abutting pair of features that appear to be unresolved at their interface.  (a) The lower set of contours represent the HI brightness at $-$102 km/s (2 km/s bandwidth) around ({\it l,b}) = (175\arcdeg,56\arcdeg).  The upper set of closed contours represent the {\it ILC} data.  (b) This is a schematic cross-section along an axis joining the relevant HI and {\it ILC} peaks plotted in (a).  The data were read off at small intervals along the axis marked in (a) and the scale may be judged from the fact that the HI and {\it ILC} peaks are located 1.\arcdeg2 apart.  At the interface between the two features, the edges are unresolved (see text) and notice the close agreement in overall morphology of the two forms of radiation.
}
\end{figure}
\clearpage

\begin{figure}
\figurenum{15}
\epsscale{.7}
\plotone{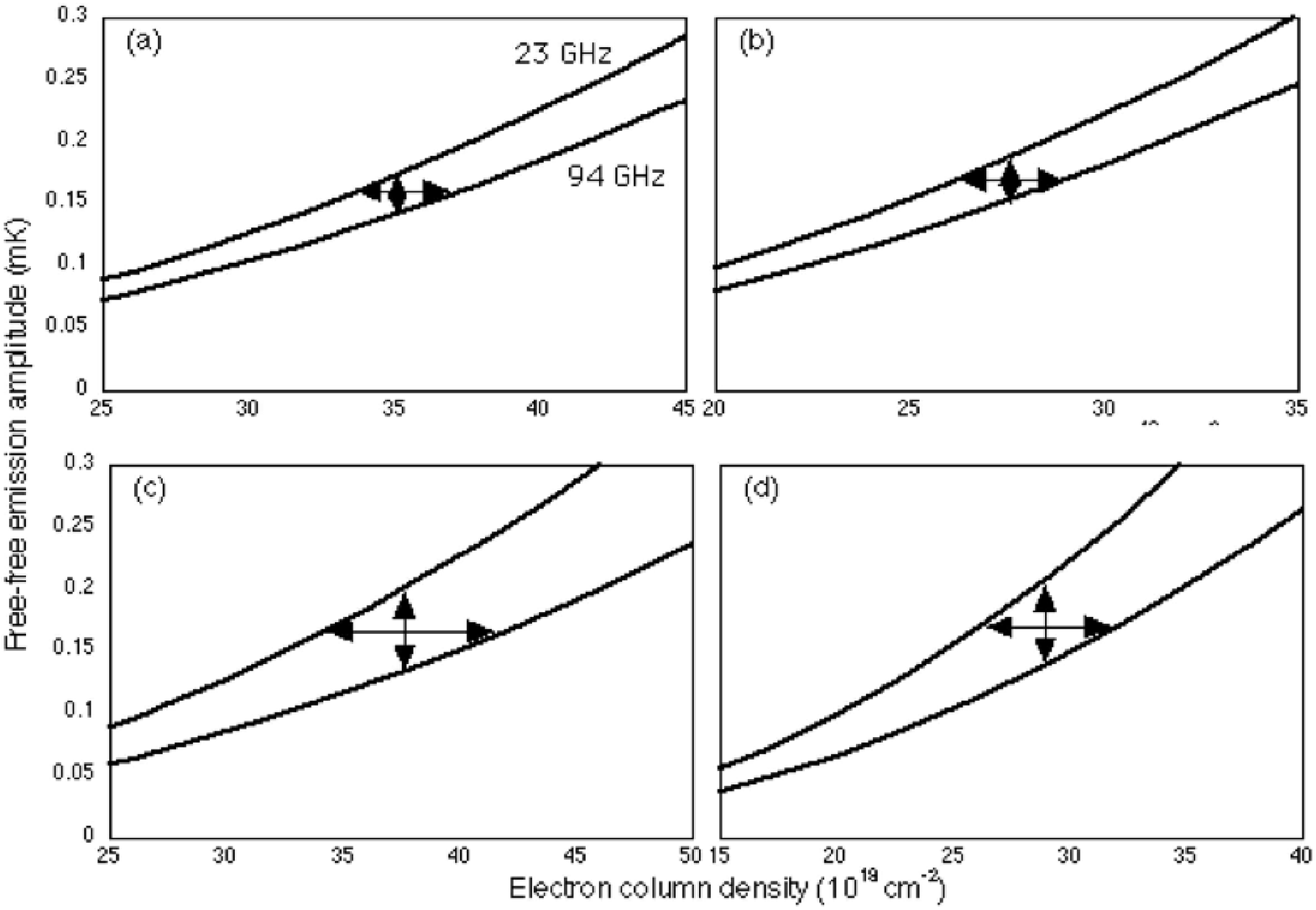}
\caption {Plots of the expected amplitude of free-free emission that would be observed at the two extremes of the {\it WMAP} frequency band from electron concentrations in interstellar space as a function of the column density of the electrons in units of 10$^{18}$ cm$^{-2}$.  (a) Model 1: Distance 35 pc for electron temperature 8,000 K in a feature 6 \arcmin\ in diameter and Aspect Ratio $=$ 1.  The horizontal line indicates the average value of the observed (positive) signal in the {\it ILC} data.  (b) Model 3: Distance 100 pc for an electron temperature of 8,000 K for a feature 6\arcmin\ in diameter and Aspect Ratio $=$ 0.2.  (c) Model 2: Distance 35 pc, electron temperature 50K in a feature 1\arcmin\ in diameter and Aspect Ratio $=$ 1 . (d) Model 4: Distance 100 pc, electron temperature 50K in a feature 1\arcmin\ in diameter and Aspect Ratio $=$ 0.2. See text.}
\end{figure}
\end{document}